\documentclass[fleqn,usenatbib]{mnras}

\usepackage{newtxtext,newtxmath}
\usepackage[T1]{fontenc}

\DeclareRobustCommand{\VAN}[3]{#2}
\let\VANthebibliography\thebibliography
\def\thebibliography{\DeclareRobustCommand{\VAN}[3]{##3}\VANthebibliography}

\usepackage{graphicx}	% Including figure files
\usepackage{amsmath}	% Advanced maths commands

\title[Ages, metallicities, and $UVJ$ colours]{Ages and metallicities of quiescent galaxies: confronting broad-band ($UVJ$) colours with stellar absorption lines} 

\author[C. M. Cheng et al.]{
Chloe M. Cheng$^{1}$,\thanks{E-mail: cheng@strw.leidenuniv.nl (CMC)}
Mariska Kriek$^{1}$,
Aliza G. Beverage$^{2}$,
Martje Slob$^{1}$,
Rachel Bezanson$^{3}$,
Marijn Franx$^{1}$,
\newauthor
Joel Leja$^{4, 5, 6}$,
Pavel E. Mancera Piña$^{1}$,
Katherine A. Suess$^{7}$,
Arjen van der Wel$^{8, 9}$,
Jesse van de Sande$^{10}$,
\newauthor
and Pieter G. van Dokkum$^{11}$
\\
$^{1}$Leiden Observatory, Leiden University, P.O. Box 9513, NL-2300 RA Leiden, The Netherlands\\
$^{2}$Department of Astronomy, University of California, Berkeley, CA 94720, USA\\
$^{3}$Department of Physics \& Astronomy and PITT PACC, University of Pittsburgh, Pittsburgh, PA 15260, USA\\
$^{4}$Department of Astronomy \& Astrophysics, The Pennsylvania State University, University Park, PA 16802, USA\\
$^{5}$Institute for Gravitation and the Cosmos, The Pennsylvania State University, University Park, PA 16802, USA\\
$^{6}$Institute for Computational \& Data Sciences, The Pennsylvania State University, University Park, PA 16802, USA\\
$^{7}$Department for Astrophysical \& Planetary Science, University of Colorado, Boulder, CO 80309, USA\\
$^{8}$Sterrenkundig Observatorium, Universiteit Ghent, Krijgslaan 281 S9, B-9000 Gent, Belgium\\
$^{9}$Max-Planck-Institut für Astronomie, Königstuhl 17, D-69117 Heidelberg, Germany\\
$^{10}$School of Physics, University of New South Wales, NSW 2052, Australia\\
$^{11}$Astronomy Department, Yale University, 52 Hillhouse Avenue, New Haven, CT 06511, USA}

\date{Accepted 2025 May 7. Received 2025 April 23; in original form 2025 February 26}

\pubyear{\the\year{}}

\begin{document}
\label{firstpage}
\pagerange{\pageref{firstpage}--\pageref{lastpage}}
\maketitle

\begin{abstract}
For decades, studying quiescent galaxies beyond $z\sim1$ has been challenging due to the reliance on photometric spectral energy distributions, which are highly susceptible to degeneracies between age, metallicity, dust, and star-formation history.  Only recently has deep, rest-frame, optical spectroscopy made robust metallicity and age measurements possible, allowing us to empirically assess their effects on continuum shapes.  To this end, we measure ages and metallicities of $\sim700$ massive ($10.2\lesssim\log(M_*/M_\odot)\lesssim11.8$), quiescent galaxies at $0.6\lesssim z\lesssim1.0$ from the Large Early Galaxy Astrophysics Census (LEGA-C) via continuum-normalized, absorption-line spectra, and compare with independent rest-frame $U-V$ and $V-J$ colours.  Age increases along the quiescent sequence as both colours redden, consistent with stellar population synthesis (SPS) model predictions.  Metallicity increases perpendicularly to the age trend, with higher metallicities at redder $U-V$ and bluer $V-J$ colours.  Thus, age and metallicity behave differently in the $UVJ$ diagram.  Moreover, this trend conflicts with SPS model expectations of increasing metallicity approximately along the quiescent sequence.  Independent dynamical mass-to-light ratio trends also differ dramatically from SPS model predictions.  These results demonstrate that relying on model fits to continuum shapes alone may lead to systematic biases in ages, metallicities, and stellar masses.  The cause of these data-model disparities may stem from non-solar abundance patterns in quiescent galaxies or the treatment of evolved stellar phases in the models.  Resolving these discrepancies is crucial, as photometric data remain central even with \textit{JWST}.
\end{abstract}


%%%%%%%%%%%%%%%%%%%%%%%%%%%%%%%%%%%%%%%%%%%%%%%%%%

%%%%%%%%%%%%%%%%% BODY OF PAPER %%%%%%%%%%%%%%%%%%

\section{Introduction}\label{sec:introduction}
Stellar populations of massive quiescent galaxies provide insight into the physical processes driving galaxy formation and evolution.  In particular, tracing these populations over cosmic time is crucial for distinguishing different galaxy assembly and quenching mechanisms (e.g., \citealt{Matteucci_1994, Trager_2000, Conroy_2014, Choi_2014, Peng_2015, Kriek_2016, Kriek_2019, Spitoni_2017, Maiolino_Mannucci_2019, Trussler_2020, Beverage_2021, HM_Beverage, Beverage_suspense, Cheng_2024}).  These investigations subsequently provide constraints on cosmological simulations and theoretical models for galaxy formation and evolution \citep{Maraston_2006, Maiolino_Mannucci_2019}.

A common approach to understanding quiescent stellar populations is to fit broad-band spectral energy distribution (SED) shapes with stellar population synthesis (SPS) models \citep{Tinsley_1972, Searle_1973, Larson_Tinsley_1978}.  broad-band photometric data are relatively inexpensive to obtain and primarily reveal information about stellar mass-to-light ($M/L$) ratios, specific star-formation rates, and dust attenuation (e.g., \citealt{Sawicki_Yee_1998, Bell_deJong_2001,  Papovich_2001, Bell_2003, Gallazzi_Bell_2009, Wuyts_2009, Zibetti_2009, Reddy_2010, Reddy_2012, Leja_2019, Abdurrouf_2021, prospector}; see \citealt{Conroy_2013} as well).  They can also indicate ages and metallicities (e.g., \citealt{Bell_deJong_2000, MacArthur_2004, Whitaker_2010, Pacifici_2016, Nersesian_2024, Nersesian_2025}); however, these measurements are challenging due to well-known degeneracies between various stellar properties, including age, metallicity, star-formation history (SFH), and dust \citep{Worthey_1994, Bell_deJong_2001, Papovich_2001, Bruzual_2003, Gallazzi_2005, Leja_2017, Leja_2019_sfh, Leja_2019_prospector}.   

While the degeneracy between age and dust can be broken with longer wavelength data (i.e. $UVJ$ colours; \citealt{Burgarella_2005, Labbe_2005, Wuyts_2007, Leja_2017}), the age--metallicity degeneracy cannot reliably be broken with photometric data alone \citep{Worthey_1994, Lee_2007, Eminian_2008, Conroy_2013, Nersesian_2024, Nersesian_2025}.   In this case, spectroscopic data can be used to disentangle and infer these quantities robustly.  By fitting Lick indices \citep{Burstein_1984, Worthey_lick} or the full spectrum \citep{CvD_2012a, Conroy_2018} with SPS models, one can directly quantify ages, metallicities, and elemental abundances.  However, high-quality spectra are required to make these measurements.  Thus, while this has been done extensively in local quiescent galaxies (e.g., \citealt{Gonzalez_1993, Caldwell_1998, Trager_2000, Vazdekis_2001, Bruzual_2003, Gallazzi_2005, Thomas_2005, Graves_2008, Choi_2014, Conroy_2014, Gu_2018, Bernardi_2023}), these spectra were historically difficult to obtain beyond $z \sim 1$ as faint absorption features needed to measure ages and metallicities are shifted into the near-infrared.  With developments allowing for ultra-deep ground- and space-based spectra, these analyses are now also being extended to high redshifts (up to $z\sim 3$), where massive quiescent galaxies are found to be metal-poor and $\alpha$-enhanced compared to their low-$z$ counterparts (e.g., \citealt{Kriek_2016, Kriek_2019, Jafariyazani_2020, Jafariyazani_2024, Beverage_2021, Beverage_2023, HM_Beverage, Beverage_suspense, Carnall_2022, Zhuang_2023}).  

Ideally, to maximize the information being considered when modelling stellar populations, one can fit photometric and spectroscopic data simultaneously (see e.g. \citealt{Kriek_2008, Belli_2019, Tacchella_2022, Akhshik_2023, Cappellari_2023, Kaushal_2024, Slob_2024, Nersesian_2025}).  While powerful constraints on SFHs and, therefore, quenching mechanisms can be obtained by leveraging the complementary strengths of spectroscopy and photometry to break the degeneracies discussed above \citep{Tacchella_2022}, models often struggle to reproduce both types of data simultaneously (e.g., \citealt{Fumagalli_2016, Carnall_2024, Kriek_2024, Slob_2024}).  Additionally, different available SPS models can lead to different and inconsistent stellar population properties (e.g., \citealt{Lee_2007, Eminian_2008, Muzzin_2009, FSPS2, Walcher_2011, Pacifici_2023, Whitler_2023}), in particular when including redder wavelength data for quiescent galaxies \citep{van_de_sande_2015}.  These discrepancies may be due to the fact that there are a variety of uncertainties associated with SPS models, as different models make varying assumptions about stellar evolution and implement different spectral libraries, as well as to the choices that the user can make for the models and priors within the available parameter spaces (\citealt{Walcher_2011, Conroy_2013}, see also \citealt{Coelho_2020, Byrne_2023, Bellstedt_2024, Nersesian_2024, Slob_2024}).  Thus, resulting galaxy properties may be uncertain and systematically biased.  In combination with the modelling degeneracies discussed above, possible template mismatches are therefore concerning.

To overcome some of these issues and better understand the inconsistencies between both continuum shapes and absorption lines and different SPS models, we require a large sample of distant quiescent galaxies with high-quality (i) broad-band photometry and (ii) age and metallicity measurements from deep, absorption-line spectra that are independent of the shape of the continuum.  This is now possible with the Large Early Galaxy Astrophysics Census (LEGA-C; \citealt{van_der_Wel_2016, van_der_Wel_2021, Straatman_2018}), a deep spectroscopic survey of galaxies at $0.6\lesssim z \lesssim 1.0$ with UltraVISTA photometry \citep{McCracken_2012}.  In this work, we model the spectra of $\sim700$ quiescent galaxies from LEGA-C with the \textsc{absorption line fitter} (\textsc{alf}, which continuum-normalizes spectra prior to fitting, \citealt{CvD_2012a, Conroy_2018}).  We combine rest-frame $UVJ$ colours with independent, SSP-equivalent ages and metallicities to understand the relationships between these quantities as well as how our data compare to popular SPS models.

This paper is organized as follows: in Section~\ref{sec:data_sample} we summarize the LEGA-C data set and detail the full-spectrum fitting employed to derive galaxy ages and metallicities.  In Section~\ref{sec:results} we present our results.  In particular, we discuss relationships between age and colour and metallicity and colour in Section~\ref{sec:relations}.  We also compare our results to predictions from two popular SPS models, the Flexible Stellar Population Synthesis models (\textsc{fsps}; \citealt{FSPS1, FSPS2}) and the \cite{Bruzual_2003} models, in Section~\ref{sec:tracks}.  We compare our results with previous work, discuss the implications for galaxy evolution studies and SPS models, and outline some caveats of our work in Section~\ref{sec:discussion}.  Finally, we summarize and conclude in Section~\ref{sec:conclusions}.

Throughout this work, we assume a flat Lambda cold dark matter ($\Lambda$CDM) cosmology with $\Omega_m = 0.3$, $\Omega_\Lambda = 0.7$, and $H_0 = 70{\ }{\rm km}{\ }{\rm s}^{-1}{\ }{\rm Mpc}^{-1}$.  All magnitudes are given in the AB-magnitude system \citep{Oke_1983}.  A \cite{Kroupa} initial mass function is assumed throughout.

\section{Data and methods}\label{sec:data_sample}
\begin{figure*}
    \centering
    \includegraphics[width=0.8\textwidth]{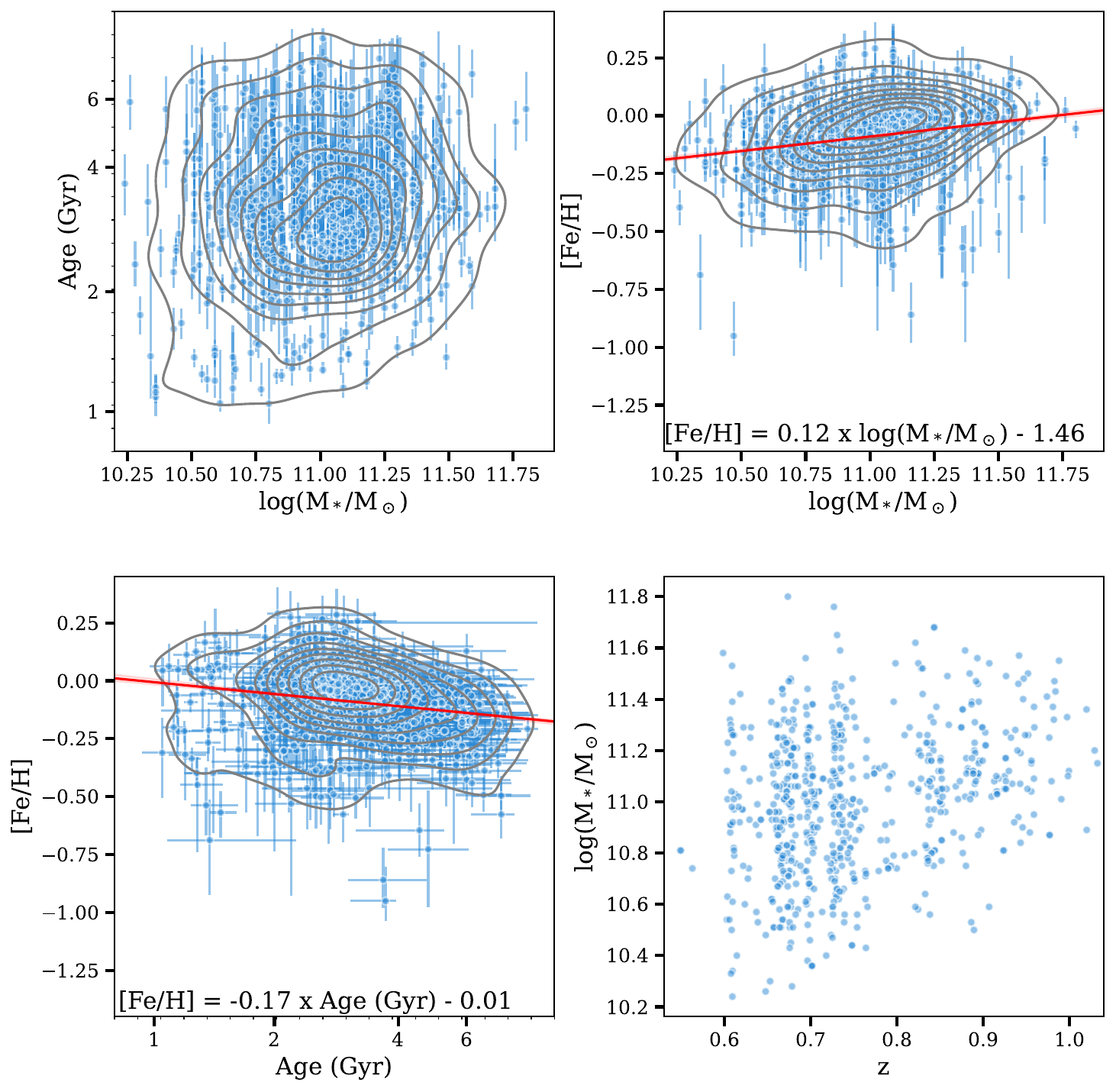}
    \caption{Properties of our final sample of massive quiescent galaxies from LEGA-C.  In the top left panel, we show SSP-equivalent age as a function of stellar mass.  In the top right panel, we show [Fe/H] as a function of stellar mass.  In the bottom left panel, we show [Fe/H] as a function of age.  In the bottom right panel, we show stellar mass as a function of redshift.  We fit linear relations to the points in the top right and bottom left panels, respectively (solid lines with shaded uncertainties) and we state the corresponding best-fitting relations at the bottom of these panels.  Stellar masses are measured by the LEGA-C team using \textsc{magphys} \citep{MAGPHYS} and SSP-equivalent ages and [Fe/H] abundances are derived from our \textsc{alf} fits.}
    \label{fig:sample_properties}
\end{figure*}

In this work, we use data from LEGA-C \citep{van_der_Wel_2021}.  We provide a brief summary of the survey and sample selection in Section~\ref{sec:data} and direct the reader to \cite{van_der_Wel_2021} and \cite{Cheng_2024} (C24 hereafter) for more details.  In Section~\ref{sec:alf}, we summarize the methods used to derive stellar population ages and elemental abundances for each data set.  

\subsection{LEGA-C}\label{sec:data}
We use spectroscopic data from the third data release of LEGA-C, a European Southern Observatory (ESO) Public Spectroscopic survey of 3600 galaxies at $0.6 \lesssim z \lesssim 1.0$.  Galaxies were selected from the UltraVISTA \citep{McCracken_2012} $K$-band catalogue by \cite{Muzzin_2013} and are located in the Cosmic Evolution Survey (COSMOS) field \citep{Scoville_2007}.  The data were collected using the VIsible MultiObject Spectrograph (VIMOS) on the ESO Very Large Telescope (VLT) over 128 nights, providing deep (20-h integration), $R\sim3500$ spectra with signal-to-noise (S/N) $\sim20$ \AA$^{-1}$ on average.  The raw 2D and reduced 1D spectra are available on the ESO Science Archive Facility\footnote{\url{http://archive.eso.org/eso/eso_archive_main.html}~.}\textsuperscript{,}\footnote{The reduced 1D spectra and catalogue have been released by ESO (\url{http://archive.eso.org/cms/eso-archive-news/Third-and-final-release-of-the-Large-Early-Galaxy-Census\\-LEGA-C-Spectroscopic-Public-Survey-published.html}).  They are also available here: \url{https://users.ugent.be/~avdrwel/research.html\#legac}~.}.  See \cite{van_der_Wel_2016, van_der_Wel_2021}, and \cite{Straatman_2018} for details.  

We use the LEGA-C sample selected in C24.  We summarize the sample selection and stellar population parameter derivation here, but refer the reader to C24 for details.  In particular, C24 determined rest-frame $UVJ$ colours using \textsc{EaZY} \citep{EAZY}, via the \cite{Muzzin_2013} $K$-band catalogue\footnote{There is a known systematic offset in the \cite{Muzzin_2013} photometry due to zero-point corrections \citep{van_der_Wel_2021}.  This offset is corrected in the UltraVISTA DR3 catalogue which does not cover the entire LEGA-C field, and thus we use the original \cite{Muzzin_2013} catalogue.  We test our analysis for the common galaxies ($\sim 290$) in the two catalogues.  The redetermined $UVJ$ colours are consistent between both sets of photometry, and the use of the DR3 photometry does not change our conclusions.} with the spectroscopic redshifts fixed to those measured from the LEGA-C spectra \citep{van_der_Wel_2021}, and selected quiescent galaxies by employing the classification from \cite{Muzzin_2013_uvj}.  C24 also removed galaxies with spectroscopic redshifts $< 0.5$.  Additionally, galaxies in LEGA-C's mask 2 were discarded as their noise spectra are significantly underestimated.  A median rest-frame S/N $\gtrsim20$ \AA$^{-1}$ was required, as well as a maximum wavelength of at least $4450$ \AA\ in order to sufficiently recover ages and metallicities.  Note that the sample size in C24 was limited by the S/N of the spatially resolved spectra.  However, this study is concerned with the full, integrated spectra.  As such, we are able to add 244 more quiescent galaxies with S/N $\gtrsim20$ \AA$^{-1}$.

\subsection{Spectral extraction and full-spectrum modelling}\label{sec:alf}
In this paper, we use the 1D spectra from C24, which were extracted from the 2D LEGA-C spectra as part of a spatially resolved analysis.  In summary, an optimal extraction procedure was performed to obtain spatially resolved and integrated spectra, by fitting the spectral flux profiles with \cite{Moffat_1969} profiles.  To extract the integrated spectra, these fits were used to perform weighted sums of all rows with significant flux (i.e. excluding (sub-)rows outside of the 3rd and 97th percentiles of the Moffat profiles to reduce noise, with the sums weighted by the Moffat profiles).  For consistency, we perform the same extraction of the integrated spectra for the additional 244 galaxies in our sample.  See C24 for details.

C24 derived stellar population parameters from the LEGA-C spectra using the \textsc{absorption line fitter} (\textsc{alf}\footnote{\url{https://github.com/cconroy20/alf}}; \citealt{CvD_2012a, Conroy_2018}).  Empirical simple stellar populations (SSPs) constructed using the Mesa Isochrones and Stellar Tracks (MIST; \citealt{MIST}) and the Spectral Polynomial Interpolator  (SPI; \citealt{Villaume_2017})\footnote{\url{https://github.com/AlexaVillaume/SPI_Utils}~.} provide the foundation for the \textsc{alf} models.  \textsc{alf} employs the Medium Resolution INT Library Of Empirical Spectra (MILES; \citealt{Sanchez_Blazquez_2006}), the Extended Infrared Telescope Facility stellar library (E-IRTF; \citealt{Villaume_2017}), and a large sample of M-dwarf spectra \citep{Mann_2015} with SPI to generate stellar spectra as a function of effective temperature ($T_{\mathrm{eff}}$), surface gravity, and metallicity from a data-driven model.

The empirical parameter space is set by the combination of the E-IRTF and \cite{Mann_2015} samples.  It spans $-2.0\lesssim$ [Fe/H] $\lesssim 0.5$ and $3.5\lesssim\log{(T_{\mathrm{eff}}/{\rm K})}\lesssim 3.9$.  To ensure high-quality interpolation at the boundaries of the empirical parameter space, \textsc{alf} also makes use of a theoretical stellar library (C3K; see \citealt{CvD_2012a}).  By differentially including theoretical element response functions, \textsc{alf} allows for variable abundance patterns.  The elemental abundances of the base stellar population are estimated in \cite{Milone_2011} and used in conjunction with the differential response functions to estimate the absolute abundances. 

During the fitting procedure, \textsc{alf} removes the continuum from the observations by fitting a high-order Chebyshev polynomial to the ratio of the data to the model.  It then implements a Fortran version of the Markov chain Monte Carlo (MCMC) algorithm \texttt{emcee} \citep{emcee} to sample the posteriors of 46 stellar population parameters.  In particular, it allows for arbitrary variation in stellar population age and detailed elemental abundance patterns.  To characterize observed errors, \textsc{alf} also fits for systematic parameters.  Note that \textsc{alf} fits spectra between $3700$ and $24000$ \AA\ and stellar populations older than 1 Gyr.  For details, see \cite{CvD_2012a} and \cite{Conroy_2018}.  

C24 fit each spectrum with 1024 walkers, 20000 burn-in steps, and a 1000-step production run.  For each fit, 500 MCMC chains were examined.  A \cite{Kroupa} initial mass function (IMF) was assumed and a single stellar population age was fit.  To initialize the age of each galaxy and to avoid the \textsc{alf} walkers getting trapped at an unrealistically high initial age, a random value was drawn from a uniform distribution centred at 3 Gyr.  Additionally, the upper limit of the age prior was set to be the age of the Universe at the redshift of each galaxy, plus 2 Gyr to allow for uncertainties.  The hot star component was not fit.  Variation in all stellar parameters in \textsc{alf} was allowed.  Galaxies for which \textsc{alf} fit an age $< 1$ Gyr were discarded (5 galaxies), as the abundances are extrapolated for ages younger than 1 Gyr and are unreliable \citep{Conroy_2018}.  Furthermore, the fits for galaxies where the posteriors ran up against the age, [Z/H], and [Fe/H] priors were discarded as these fits would result in unreliable ages and metallicities ($\sim 6$ per cent of the sample).  There are a variety of reasons why the parameter space may not be fully explored for these objects, for example low S/N, low levels of residual star formation, or non-stellar absorption.  We fit the 244 galaxies that we add to this sample in exactly the same way as C24.  Finally, we visually inspect all of the spectra and remove any objects near the edge of the $UVJ$ box with strong [O$_{\rm III}$] emission (10 galaxies), which could potentially host active galactic nuclei or have low levels of star formation. The final LEGA-C sample examined in this study includes 683 quiescent galaxies.

We show the properties of our final sample in Fig.~\ref{fig:sample_properties}.  In the top panels, we show, respectively, SSP-equivalent ages and [Fe/H] abundances versus stellar masses (computed using \textsc{magphys}, \citealt{MAGPHYS}, and obtained via private communication with the LEGA-C team).  In the bottom-left panel, we show our SSP-equivalent ages versus SSP-equivalent [Fe/H] abundances.  In the bottom right panel, we show stellar masses versus spectroscopic redshifts.  We remove two galaxies with \textsc{magphys} masses $< 10^{10}{\ }{\rm M}_\odot$ as the flux levels of their spectra are abnormally low.  Overall, the galaxies in our sample are at $0.6 \lesssim z \lesssim 1.0$ with $10.2 \lesssim \log(M_*/M_\odot)\lesssim 11.8$ and $-1.3\lesssim {\rm [Fe/H]}\lesssim 0.3$, and are $1.0 - 8.6$ Gyr old. 

A subset of 135 quiescent LEGA-C galaxies were fit using \textsc{alf} in \cite{Beverage_2023}.  They fit the 1D extracted spectra from the LEGA-C public release.  We find that the stellar population parameters for the galaxies that we have in common (110 galaxies) are fully consistent between the two studies, despite subtle differences in the spectral extraction procedures (not shown).  We also qualitatively recover the negative relation between [Fe/H] and age \citep{Beverage_2021} and the slightly positive relation between [Fe/H] and mass \citep{Beverage_2023}, as shown by the red lines in Fig.~\ref{fig:sample_properties}.  We state the linear fits in the corresponding panels in Fig.~\ref{fig:sample_properties}. 

\begin{figure*}
    \centering
    \includegraphics[width=\textwidth]{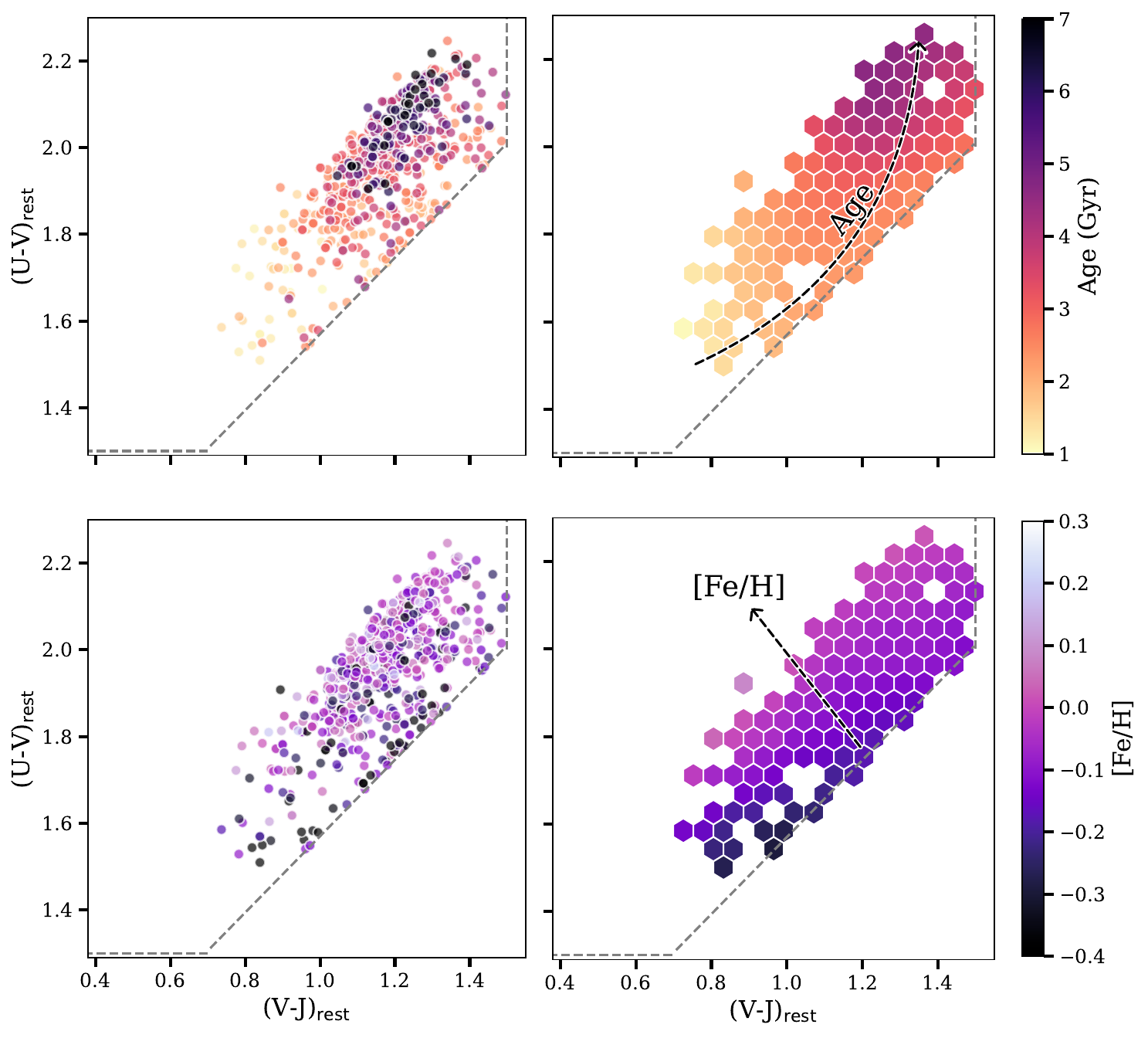}
    \caption{Rest-frame $UVJ$ diagrams of quiescent galaxies at $0.6\lesssim z\lesssim 1.0$ from the LEGA-C survey.  \textit{Left}: In these panels, we colour code the symbols by their measured ages (top) and their measured [Fe/H] values (bottom), as derived by fitting absorption lines in continuum-normalized spectra with \textsc{alf}.  \textit{Right}: In these panels, we LOESS-smooth the ages and metallicities from the left panels \citep{LOESS, LOESS_2D}.  In each panel, we plot an arrow showing the general, qualitative direction of the dependence between age and colour (top) and metallicity and colour (bottom).}
    \label{fig:UVJ_figure}
\end{figure*}

\section{Results}\label{sec:results}
In this paper we compare detailed stellar population properties derived from absorption lines to colours derived from broad SED shapes.  In Section~\ref{sec:relations}, we explore how age and metallicity vary on the $UVJ$ diagram.  In Section~\ref{sec:tracks}, we compare our results with predictions from popular SPS models.

\subsection{Trends with age and metallicity}\label{sec:relations}
In Fig.~\ref{fig:UVJ_figure}, we show several rest-frame $UVJ$ diagrams colour coded by age and metallicity.  In the left panels, we present the results for the individual galaxies in our LEGA-C sample.  The symbols are colour coded by their measured SSP-equivalent ages (top) and [Fe/H] values (bottom), both measured via absorption lines in continuum-normalized spectra (Section~\ref{sec:alf}).  In the right panels, we smooth the data using the Locally Weighted Regression (LOESS) 2D algorithm \citep{LOESS_2D} via the \textsc{loess} Python package \citep{LOESS}.  

There is a strong sequence in age in the top panels.  In particular, age increases approximately parallel to the quiescent sequence, from $\sim 1$ Gyr in the bluest regions of the $UVJ$ diagram to $\sim 3-5$ Gyr in the reddest regions.  There is some curvature to this trend, with the oldest galaxies populating the reddest end of the $U-V$ axis, but located slightly blueward on the $V-J$ axis.  Interestingly, [Fe/H] behaves differently, increasing approximately perpendicular to the quiescent sequence, from subsolar ($-0.3 \lesssim $ [Fe/H] $\lesssim -0.2$) near the edge of the $UVJ$ box to supersolar ([Fe/H] $\sim 0.08$) as one moves across the quiescent sequence.  We indicate the approximate, qualitative trends that we describe here with arrows in the right panels.

This observed age trend has been suggested in previous work (e.g., \citealt{Whitaker_2012, Whitaker_2013, Mendel_2015, Belli_2019, Diaz_Garcia_2019}).  However, in these studies, galaxy ages were derived either via the broad-band photometric SEDs alone, or by including the SEDs in simultaneous fits to spectra and photometry.  In contrast, our ages are completely independent of continuum shape as they are derived directly from the continuum-normalized LEGA-C spectra.  

\begin{figure*}
    \centering
    \includegraphics[width=\textwidth]{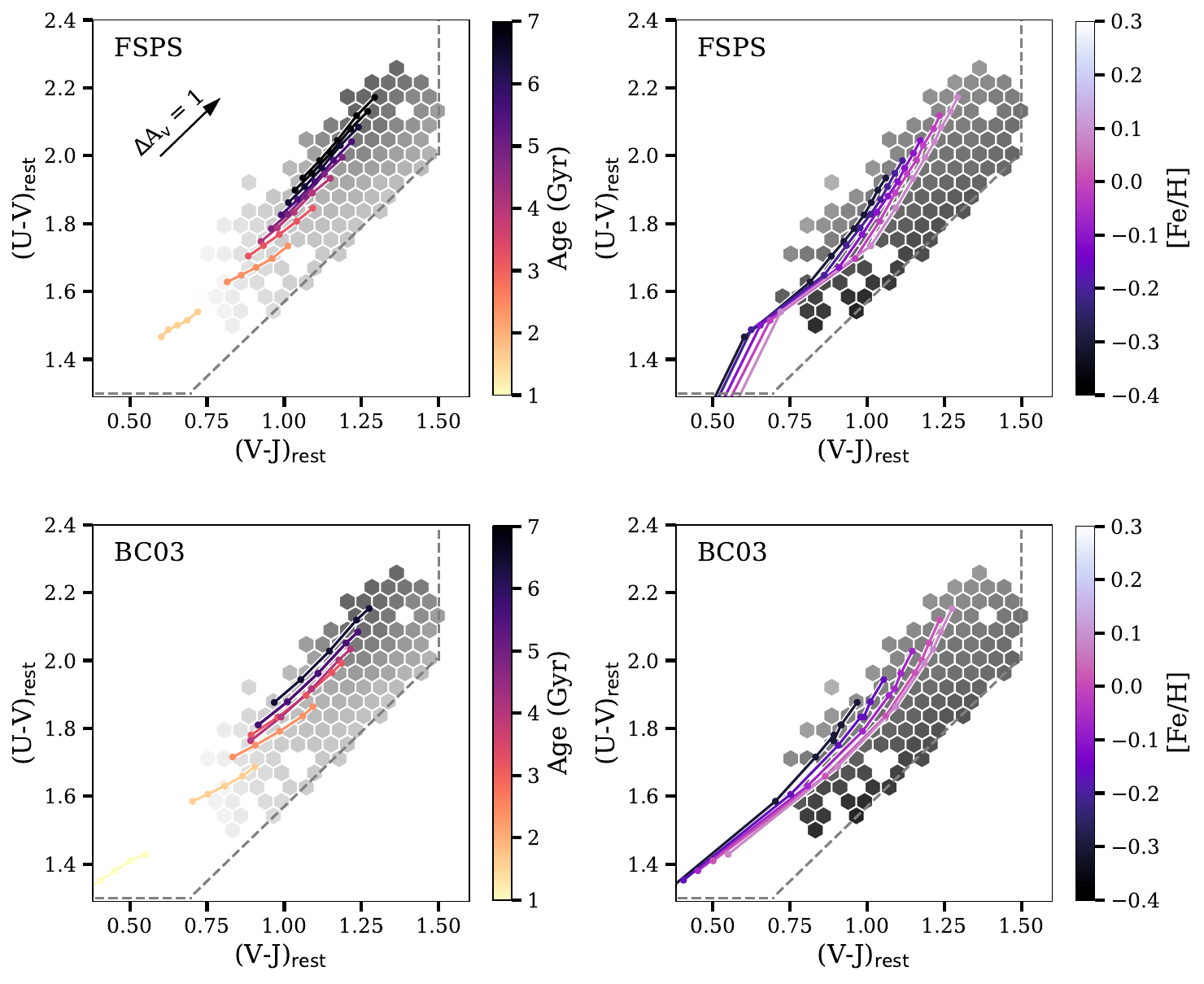}
    \caption{The same LOESS-smoothed rest-frame $UVJ$ diagrams as in the right panels of Fig.~\ref{fig:UVJ_figure} (grey) but with stellar evolution tracks overplotted.  Tracks produced using \textsc{python-fsps} (\citealt{python-FSPS}, see also \citealt{FSPS1, FSPS2}) are shown in the top panels and tracks produced using \textsc{bagpipes} (\citealt{bagpipes}, which uses the \citealt{Bruzual_2003}, or BC03, SPS models) are shown in the bottom panels.  For each model, we produce tracks spanning the age ($\sim1-8$ Gyr) and [Fe/H] ($-0.31 < $ [Fe/H]/dex $< 0.08$) ranges of our sample.  The tracks in the left and the right panels are the same tracks for each model.  On the left, we emphasize the age trend and the ages are indicated by colour-coded points, connected by lines to guide the eye.  On the right, we emphasize the [Fe/H] trend, with metallicities indicated by colour-coded points connected by lines.  In general, we see the same kind of sequence in age in the stellar evolution tracks from both models that we see in our data (left panels).  On the other hand, the metallicity sequence of the tracks seems to be perpendicular to that observed in the LEGA-C data (right panels).}
    \label{fig:loess_tracks}
\end{figure*}

Moreover, for the first time, we observe a trend in metallicity as derived from absorption lines.  While ages and metallicities are known to be degenerate \citep{Worthey_1994, Bruzual_2003, Gallazzi_2005}, previous studies that examined the relationship between ages and broad-band colours (e.g., \citealt{Whitaker_2012, Whitaker_2013, Mendel_2015, Belli_2019, Diaz_Garcia_2019}) did not take metallicities into account\footnote{Note that \cite{Diaz_Garcia_2019} examined trends between $UVJ$ colours and metallicity, however, they derived total metallicity from photometry alone and used a different SPS model.} as they were poorly constrained for distant quiescent galaxies.  However, it is now possible to measure robust metallicities beyond the local Universe and disentangle them from ages using a combination of deep LEGA-C spectra and \textsc{alf} \citep{CvD_2012b, Choi_2014, Beverage_2021, Beverage_2023, Gu_2022, Cheng_2024}.  As our measured age and [Fe/H] trends are distinct, this demonstrates that observed reddening along the quiescent sequence is indeed primarily due to an increase in age, as previously suggested (e.g., \citealt{Whitaker_2012, Whitaker_2013, Mendel_2015, Belli_2019, Diaz_Garcia_2019}).  Furthermore, the differing trends may indicate that age and metallicity are not strongly degenerate in this $UVJ$ space.  However, this does not imply that we can simply derive metallicities and ages directly from $UVJ$ colours as there are many other fitting degeneracies that must be considered.  Thus, we do not attempt to quantify the strength of the age and metallicity trends here, but simply state that qualitative trends exist.  We discuss this further in Section~\ref{sec:literature_comparison}.

\begin{figure*}
    \centering
    \includegraphics[width=\textwidth]{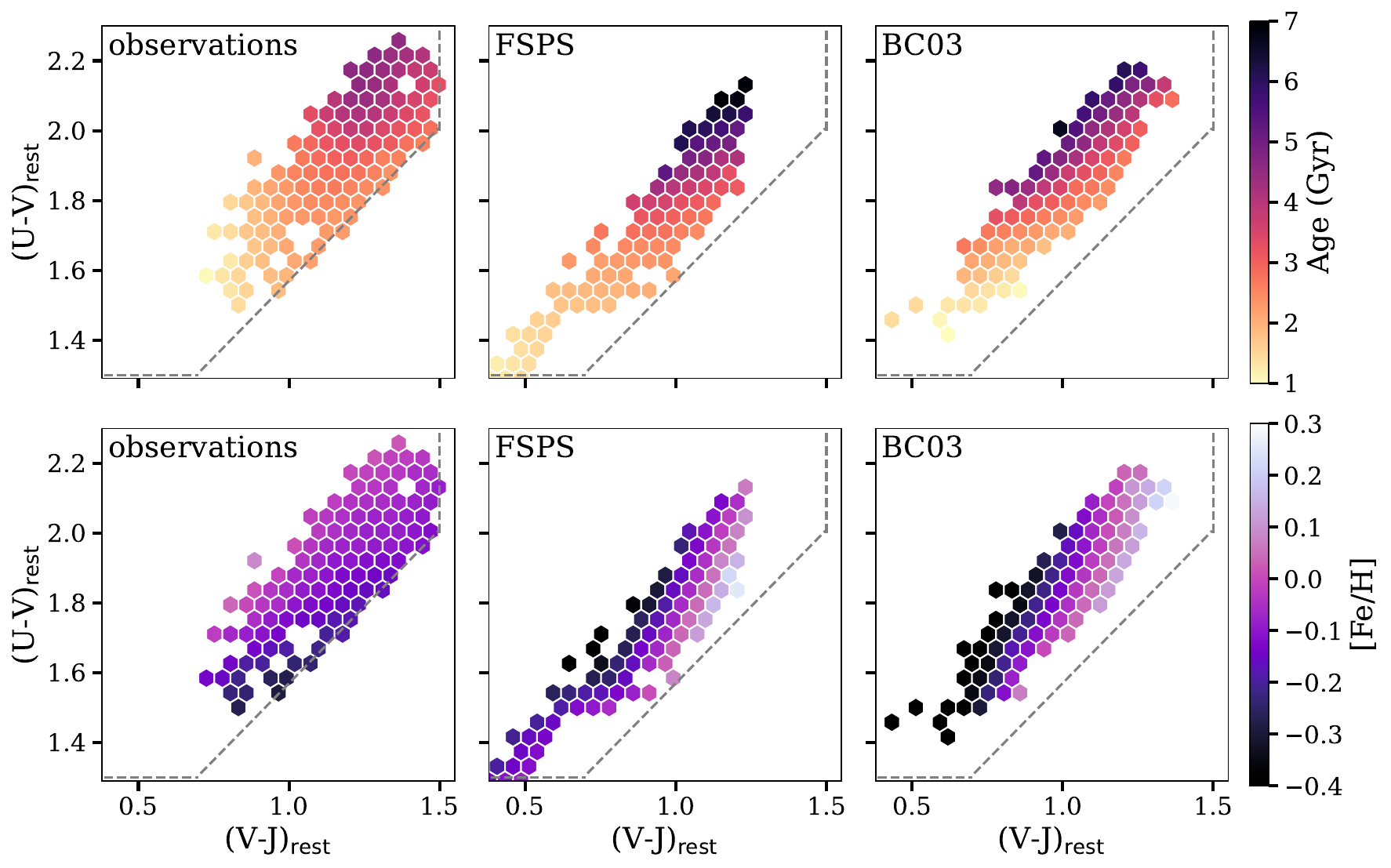}
    \caption{LOESS-smoothed rest-frame $UVJ$ diagrams for the observations and the models.  For the models, we input the ages and metallicities of each galaxy in our LEGA-C sample and output the predicted rest-frame $UVJ$ colours, with some random scatter added to simulate observational uncertainties.  The diagrams are colour coded by age in the top panels and by [Fe/H] in the bottom panels.  The same diagrams as in the right panels of Fig.~\ref{fig:UVJ_figure} are shown for the data in the left panels, the diagrams for the \textsc{fsps} models are shown in the middle panels, and the diagrams for the BC03 models are shown in the right panels.  Age follows the quiescent sequence in all three panels.  However, the trend in [Fe/H] is inconsistent between the data and the models.}
    \label{fig:UVJ_model_test}
\end{figure*}

\begin{figure*}
    \centering
    \includegraphics[width=\textwidth]{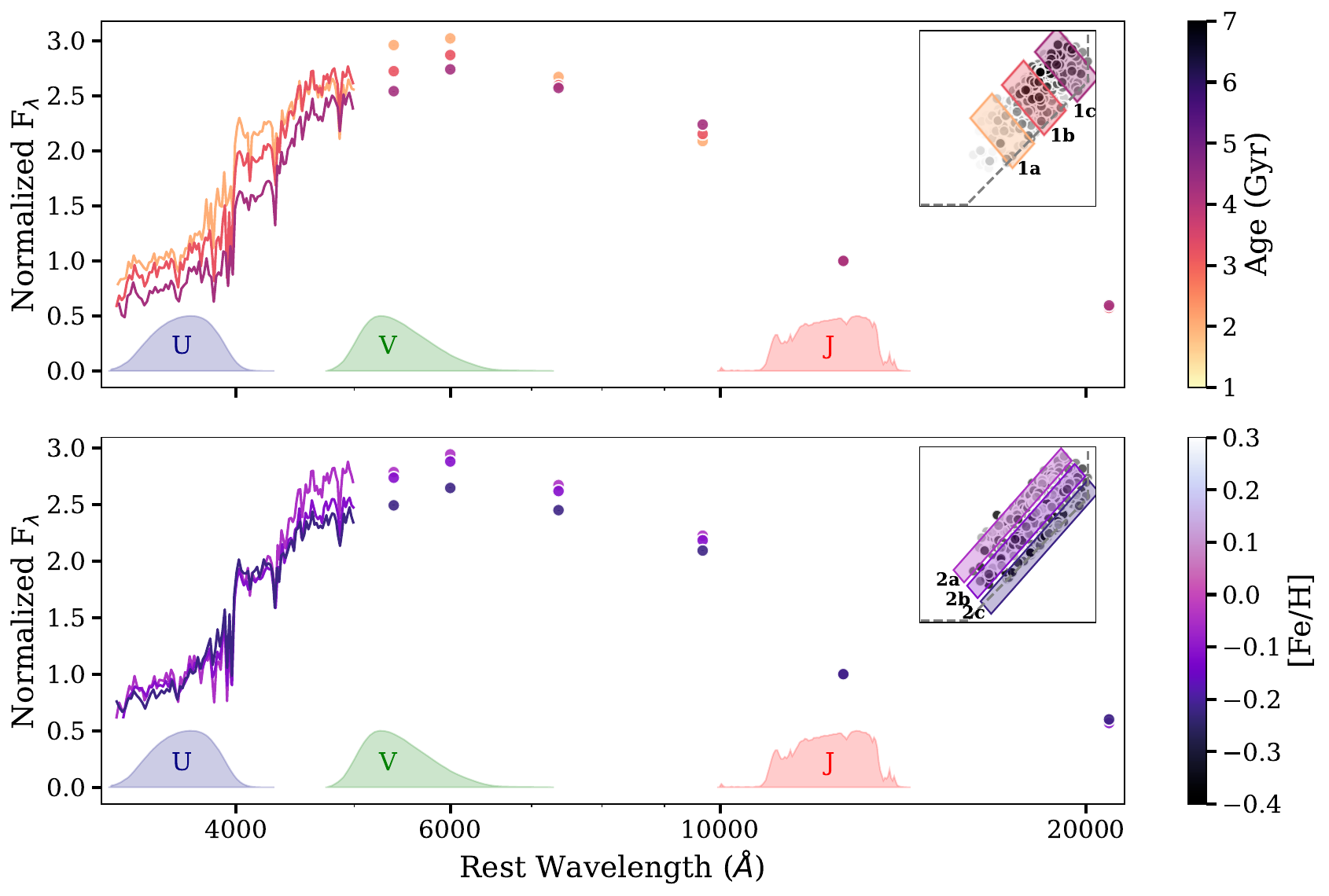}
    \caption{A comparison of stacked LEGA-C spectra.  In the top panel, we select three bins along the quiescent sequence and approximately along our observed age sequence.  In the bottom panel, we select three bins perpendicular to the quiescent sequence and approximately along our observed metallicity sequence.  We normalize the spectra over $4400-4500$ \AA\ and mean-stack the spectra in each bin.  We renormalize each stack by the photometric rest-frame $J$ band.  We also normalize and mean-stack the UltraVISTA photometry beyond $5000$ \AA\ for the same galaxies in each bin in the same way.  We colour code the stacked spectra by the median age or metallicity in each bin and median-bin the spectra in bins of 50 pixels for visibility.  We state each curve's median age, [Fe/H], sSFR averaged over the last 0.01 Gyr (measured via \textsc{magphys} by the LEGA-C team), and the number of galaxies in each stack in Table~\ref{tab:stacked_spec}.  The $UVJ$ filter response curves are indicated in each panel.  In the inset panels, we show the same rest-frame $UVJ$ diagrams as in the left panels of Fig.~\ref{fig:UVJ_figure} but greyscaled, with the bins indicated by the shaded regions.  This figure serves to illustrate our results, as in the top panel the difference in ages in the three stacked spectra (along the observed age sequence) is greater than in the bottom panel (along the observed [Fe/H] sequence), while the opposite is true for the difference in [Fe/H].}
    \label{fig:stacked_spec}
\end{figure*}

\subsection{Comparison to stellar population models}\label{sec:tracks}
Our current understanding of stellar populations in unresolved systems from both broad-band photometry and high-quality spectroscopy primarily relies on SPS models \citep{Tinsley_1972, Searle_1973, Larson_Tinsley_1978}.  Building these models involves combining isochrones, spectral libraries, and the IMF to construct simple stellar populations (SSPs).  Every model implements these components in slightly different ways.  Additionally, there are several aspects of population synthesis that still contain uncertainties \citep{Renzini_2006, Walcher_2011, Conroy_2013}.  For example, all models must make approximations for processes such as convection, stellar rotation, mass loss, binary stars, and advanced stages of stellar evolution (e.g. thermally pulsating asymptotic giant branch, or TP-AGB, stars).  Many codes treat these processes differently (compare the models of \citealt{Bruzual_2003} with those of \citealt{Maraston_2005}, for instance, and see \citealt{Muzzin_2009}).  Additionally, spectral libraries are often incomplete \citep{Coelho_2020, Byrne_2023, Bellstedt_2024}.  Thus, different SPS models can produce inconsistent results (see, e.g., \citealt{Lee_2007, Eminian_2008, Muzzin_2009, FSPS2, Walcher_2011, van_de_sande_2015, Coelho_2020, Byrne_2023}).  Moreover, they are often not able to reliably reproduce observations, either from photometry, spectra, or combined photometric and spectroscopic data, and in cases where models can reproduce the observations, the physics implied by each model can be conflicting (e.g., \citealt{Lee_2007, Fumagalli_2016, Byrne_2023, Cappellari_2023, Pacifici_2023, Kaushal_2024, Nersesian_2024, Slob_2024, Tortorelli_2024, Bevacqua_2025, Kim_2024}).

We note that several studies have fit the LEGA-C sample with different models (e.g., \citealt{Chauke_2018, Beverage_2021, Beverage_2023, Barone_2022, Cappellari_2023, Kaushal_2024, Nersesian_2024}).  However, in this work, we have access to both photometric colours from broad-band SEDs and completely independent measurements of ages and metallicities from continuum-normalized, absorption-line spectra.  Thus, we are in a position to compare our results to predictions from popular SPS models to better understand the uncertainties and discrepancies we discussed above.  To facilitate this comparison, in Fig.~\ref{fig:loess_tracks} we overplot stellar evolution tracks generated using SPS models on the same LOESS-smoothed $UVJ$-diagrams as in Fig.~\ref{fig:UVJ_figure} but with the hexbins greyscaled.  In the top panels, we show tracks computed using \textsc{python-fsps} (\citealt{python-FSPS}, and see also \citealt{FSPS1, FSPS2}), implemented with the MIST isochrones \citep{MIST, MIST1} and MILES spectral library \citep{Sanchez_Blazquez_2006, MILES2}.  In the bottom panels, we similarly show tracks computed using \textsc{bagpipes} (\citealt{bagpipes}), which makes use of the 2016 version of the \cite{Bruzual_2003} SPS models (BC03 hereafter), implemented with the Padova isochrones \citep{Padova1, Padova2, Padova3, Padova4, Padova5} and MILES spectral library.  In both cases, we model a stellar population with a $\tau$ SFH truncated after 1 Gyr and no dust (i.e., $A_V = 0$, see Section~\ref{sec:caveats} for a discussion of the impacts of our assumed SFH and $A_V$).  We indicate with an arrow the direction in which dust would move the tracks according to the \cite{Calzetti_2000} attenuation law.  From the models, we compute $UVJ$ colours for a grid of ages and metallicities, spanning $\sim1-8$ Gyr (the age of the Universe at $z\sim0.6$) and the metallicity range of our sample ($-0.31 < $ [Fe/H]/dex $< 0.08$).  We colour code the tracks by stellar population age in the left panels, where we connect points of the same age by lines to guide the eye.  We colour code the tracks by metallicity in the right panels, where we assume [Fe/H] $\approx \log\left(Z/Z_\odot\right)$ (see \citealt{Beverage_suspense} and Section~\ref{sec:model_implications}). 

In the left panels of Fig.~\ref{fig:loess_tracks}, we see that both models have age trends that are qualitatively similar to those seen in our observations.  In particular, ages generally increase along the quiescent sequence, with the trend steepening at the reddest tip of the quiescent sequence.  This comparison indicates that SPS models predict similar $UVJ$ colours as a function of age to what we observe for the LEGA-C galaxies.

However, the same cannot be said when we examine the metallicity panels (right panels of Fig.~\ref{fig:loess_tracks}).  Here, both models do not follow the trends that we see in our observations, and perhaps even show a sequence in the perpendicular direction (i.e. while our metallicities increase approximately perpendicular to the quiescent sequence, the metallicities of the models seem to increase more along the quiescent sequence).  This finding suggests that models like \textsc{fsps} and BC03 may not accurately predict broad-band colours for varying metallicities.  

We note that the model tracks in Fig.~\ref{fig:loess_tracks} only span a small range in the $UVJ$ quiescent box and that the real galaxies (grey hexagons) show a much broader distribution.  This is likely due to photometric uncertainties, differences between the real SFH and the assumed $\tau$-model SFH, differences between the real dust properties and the assumed dust extinction ($A_V = 0$), or template mismatches.  To assess the significance of our results and evaluate whether we expect to be able to observe the metallicity trend that we see in the $UVJ$ diagram, we generate the predicted $UVJ$ colours for each galaxy in our sample assuming the \textsc{fsps} and BC03 models.  In particular, we input the measured SSP-equivalent ages and metallicities of our sample and output rest-frame $UVJ$ colours for each galaxy.  Uncertainties on the rest-frame fluxes are estimated using the median S/N of the observed photometric data points (between 1 and 7 bands, depending on the filter) which are close in rest-frame wavelength.  For the $U$, $V$, and $J$ bands, we find typical uncertainties of $\sim3.5$, $\sim1.0$, and $\sim1.0$ per cent, respectively.  We then add random scatter to the model fluxes and use these to calculate realistic $U$, $V$, and $J$ colours by randomly sampling from a normal distribution multiplied by each of these typical observed relative errors.  

We plot the LOESS-smoothed $UVJ$ diagrams colour coded by age and metallicity for the observations, the \textsc{fsps} models, and the BC03 models in Fig.~\ref{fig:UVJ_model_test}.  In the top panels, the age trend is qualitatively consistent between the observations and the models.  In particular, ages always increase along the quiescent sequence, with some curvature near the top right corner of the $UVJ$ box.  In the bottom panels, however, the trend in [Fe/H] is not consistent between the observations and the models, with the trend being nearly perpendicular between the two, consistent with what we see in Fig.~\ref{fig:loess_tracks}.  This test indicates that we can realistically expect to detect the trend with metallicity in our data.  Thus, our result is significant.  Additionally, even if we are underestimating the typical relative observed errors on the $UVJ$ colours as described above, we find that we are still able to recover the trends with typical uncertainties of up to 5 per cent for each of the $U$, $V$, and $J$ bands (this percentage allows the model panels to approximately display the same spread in $UVJ$ colours as the observations).  We discuss possible additional sources of scatter in Section~\ref{sec:caveats}. 

Finally, in Fig.~\ref{fig:stacked_spec}, we visualize how the observed SEDs and spectra of galaxies change along and perpendicular to the quiescent sequence, in roughly similar directions to those in which age and metallicity increase, respectively.  In the top panel, we select three bins along the quiescent sequence (i.e. approximately along the age sequence).  In the bottom panel, we select three bins perpendicular to the quiescent sequence (i.e. approximately along the metallicity sequence).  These bins are shown by the shaded rectangles on the rest-frame $UVJ$ diagrams in the inset panels.  We normalize the spectra over the region from $4400$ to $4500$ \AA\ and mean-stack the spectra in each bin.  We then normalize each stack by the UltraVISTA photometry in the rest-frame $J$ band.  We also normalize and mean-stack the UltraVISTA photometry beyond $5100$ \AA\ for the same galaxies in each bin in the same way.  In particular, as the galaxies in our sample are at similar redshifts, we mean-stack the photometry in the observed frame and de-redshift the median wavelength of each band when plotting the points in Fig.~\ref{fig:stacked_spec}.  In the top panel, we colour code the stacked spectra and photometry by the median age in each bin and in the bottom panel we colour code by the median metallicity in each bin.  We median-bin the spectra in bins of 50 pixels for ease of visualization.  The $UVJ$ filter response curves are also indicated in each panel.  We state the median age, [Fe/H], specific star-formation rate (sSFR) over 0.01 Gyr (computed using \textsc{magphys}, \citealt{MAGPHYS}, and obtained via private communication with the LEGA-C team), and the number of galaxies in each stack in Table~\ref{tab:stacked_spec}.

\begin{table}
    \centering
    \caption{Median age, [Fe/H], sSFR over 0.01 Gyr, and number of galaxies in each stack for the stacked spectra in Fig.~\ref{fig:stacked_spec}.}
    \label{tab:stacked_spec}
    \begin{tabular}{ccccc}
        \hline
        Box & Age & [Fe/H] & sSFR (0.01 Gyr) & Number of galaxies \\
        & (Gyr) & & ($M_\odot$/yr) & \\
        \hline
        1a & 2.01 & -0.14 & -10.99 & 48 \\
        1b & 3.19 & -0.09 & -11.37 & 297 \\
        1c & 4.25 & -0.07 & -11.46 & 131 \\
        \hline
        2a & 3.66 & -0.05 & -11.58 & 297 \\
        2b & 3.22 & -0.12 & -11.21 & 185 \\
        2c & 2.67 & -0.22 & -10.73 & 62 \\
        \hline
    \end{tabular}
\end{table}

\begin{figure*}
    \centering
    \includegraphics[width=0.8\textwidth]{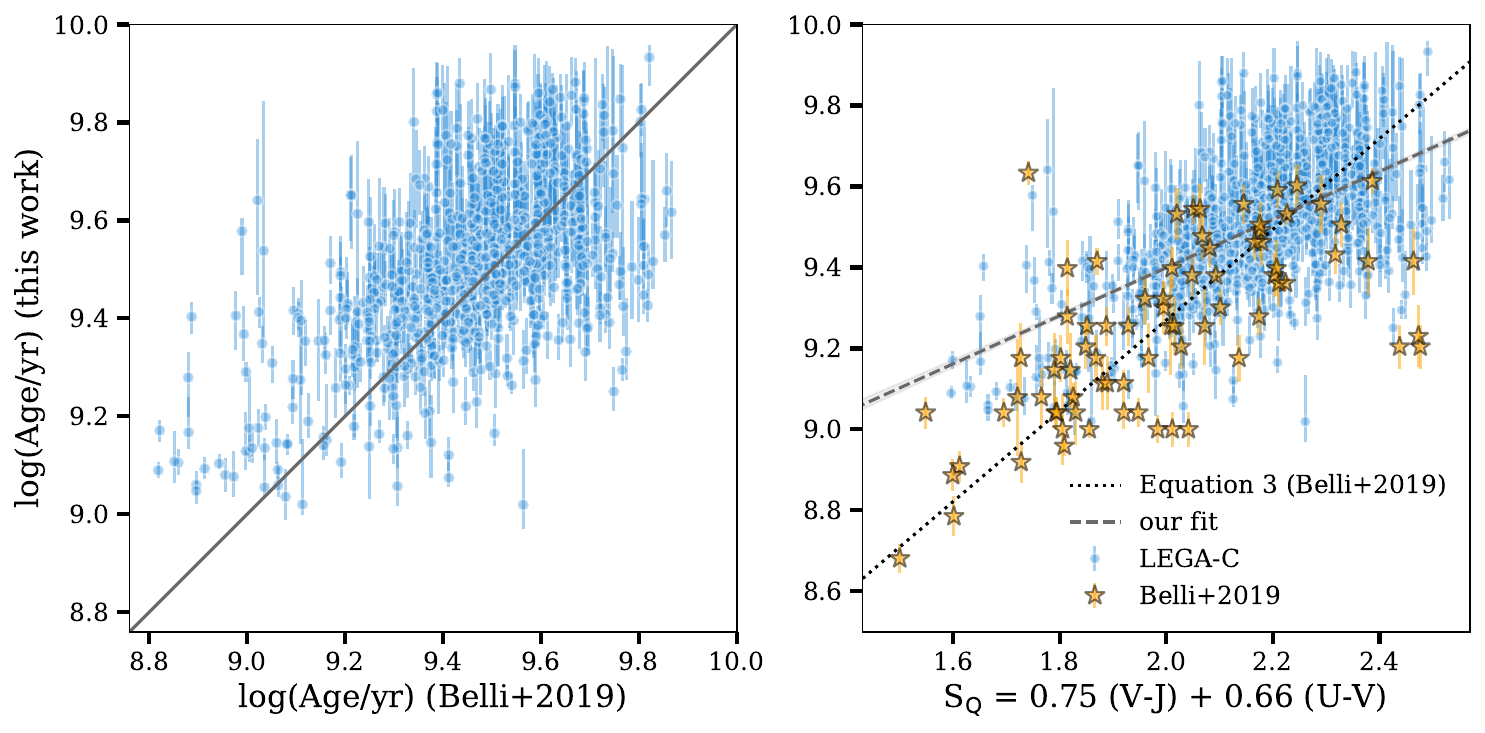}
    \caption{\textit{Left}: A comparison between the ages that we measure by using \textsc{alf} to fit absorption lines in continuum-normalized spectra ($y$-axis) to those derived using equation (3) in B19 (i.e. using only the $UVJ$ colours, $x$-axis).  The one-to-one relation is represented by the solid line.  In general, we find older ages compared to those predicted from the B19 age--colour relation.  However, overall, the ages of the LEGA-C galaxies are relatively consistent with the predicted ages from B19.  \textit{Right}: SSP-equivalent ages plotted as a function of the $S_Q$ axis defined in equation (2) in B19.  We also reproduce the points from fig. (8) in B19 (orange stars).  The dotted line is the age--colour relation from B19.  The dashed line shows our illustrative linear fit to the points in this plot, with shaded regions indicating the $1\sigma$ uncertainties.  Our fit differs qualitatively from the age--colour relation derived in B19.  Furthermore, there is a lot of scatter that is not captured by a linear relation on these axes.}
    \label{fig:Belli_compare}
\end{figure*}

This figure serves as an illustration of our results.  In particular, the difference in ages in the three stacked spectra in the top panel ($\sim 1$ Gyr on average) is greater than in the bottom panel ($\sim 0.6$ Gyr on average).  On the other hand, the difference in [Fe/H] values in the top panel ($\sim 0.05$ dex on average) is smaller than in the bottom panel ($\sim 0.1$ dex on average).  Thus, the selected bins in the top panel are tracing the age sequence that we observe with very little dependence on [Fe/H].  When we examine the combined spectra and SEDs, we see that age is primarily shifting the spectra to redder wavelengths and making the $4000$ \AA\ break more pronounced.  Meanwhile, the selected bins in the bottom panel are tracing the observed [Fe/H] sequence with a smaller dependence on age compared to the top panel.  Metallicity is not shifting the spectra in the same way as age in the top panel (i.e. the spectra all have approximately the same peak wavelength in different metallicity bins), but it acts to make the galaxies redder in $U-V$.  However, as a caveat we note that this is not a perfect comparison as metallicity and age are changing in both panels, since the observed age and metallicity trends are not exactly along and perpendicular to the quiescent sequence, respectively.  The observed shifts in the spectra here may also be affected by other degenerate effects, such as dust, SFH, or sSFR.  See Section~\ref{sec:caveats} for details. 

\section{Discussion}\label{sec:discussion}
In Section~\ref{sec:results}, we examined trends between spectroscopic ages and metallicities and independent $UVJ$ colours.  We found that observed reddening along the quiescent sequence in the $UVJ$ diagram is primarily due to age, while metallicity increases approximately perpendicular to the quiescent sequence (Section~\ref{sec:relations}).  We also investigated whether these trends can be reproduced by popular SPS models.  While these models predict similar $UVJ$ colours as a function of age compared to what we see in the data, this is not the case for metallicity (Section~\ref{sec:tracks}).  

Here, we discuss the implications of our results.  In Section~\ref{sec:literature_comparison}, we compare our results to previous studies, in particular \cite{Belli_2019}.  We discuss the implications for galaxy evolution studies in general, and for derived stellar $M/L$ ratios and masses specifically, in Section~\ref{sec:science_implications}.  We examine the implications of our results for SPS models in Section~\ref{sec:model_implications}.  Finally, in Section~\ref{sec:caveats}, we discuss caveats of our work that should be taken into account.  

\subsection{Comparison to previous work}\label{sec:literature_comparison}
In Section~\ref{sec:relations}, we used independent measurements of age and $UVJ$ colours to demonstrate that reddening along the quiescent sequence is primarily due to age, as previously suggested by observations (e.g., \citealt{Whitaker_2012, Whitaker_2013, Mendel_2015, Belli_2019, Diaz_Garcia_2019}) and cosmological simulations (e.g., \citealt{Akins_2022}).  In \cite{Belli_2019} (B19 hereafter), this trend was used to derive a relationship between ages in quiescent galaxies at $1.0 < z < 2.5$ and $UVJ$ colours (see their equation (3))\footnote{\cite{Diaz_Garcia_2019} derived similar relationships between stellar population parameters (including ages and metallicities) and $UVJ$ colours.  However, they examined photometric data only, in bands offset from $U$, $V$, and $J$.  Thus, we examine only the relationship from \cite{Belli_2019} here.}.   This relation has since been used to predict quiescent galaxy ages using only $UVJ$ colours (e.g., \citealt{Suess_2019b, Clausen_2025}).  However, in B19 and other previous studies, the broad-band photometric SEDs were included when deriving galaxy ages.  In contrast, our ages are completely independent of continuum shape as they are derived directly from the continuum-normalized LEGA-C spectra, by only fitting absorption lines using \textsc{alf}.  Thus, this study provides an independent check of the age trend along the quiescent sequence in the $UVJ$ diagram.  

In the left panel of Fig.~\ref{fig:Belli_compare}, we compare our ages measured from the LEGA-C spectra using \textsc{alf} to those derived using equation (3) in B19 (i.e. predicted from $UVJ$ colours).  In general, we find older ages compared to those predicted from the B19 age--colour relation.  This difference is somewhat surprising as we measure SSP-equivalent ages while B19 considers mass-weighted ages, and SSP-equivalent ages are typically found to be younger than mass-weighted ages \citep{Trager_2009}.  However, the difference is small, and the overall ages that we find for the LEGA-C galaxies are relatively consistent with the B19 predicted ages, even though we examine a different redshift range. Thus, the B19 relation is able to predict approximate quiescent galaxy ages down to $z\sim0.6$.  

While we confirm that the B19 relation can be used in a general sense, we find that it may be too simplistic to derive precise ages for \textit{individual} galaxies.  In the right panel of Fig.~\ref{fig:Belli_compare},  we plot our SSP-equivalent ages as a function of the $S_Q = 0.75(V - J) + 0.66(U - V)$ axis defined in B19 (see their equation (2)).  We overplot the age--colour relation derived by B19 (dotted black line, see their equation (3)) as well as their individual data points (orange stars).  As an illustrative comparison, we fit a similar linear relation to our galaxies on these axes, shown by the dashed grey line, with shaded regions indicating the $1\sigma$ uncertainty on our fit.  Our fit differs from the age--colour relation in B19.  In particular, we find a qualitatively shallower slope.  Moreover, there is a significant amount of scatter that is not captured well by either fit, likely due in part to the fact that B19 have younger galaxies (\textsc{alf} cannot fit galaxies younger than 1 Gyr, see Section~\ref{sec:alf}), and because the relationship between $UVJ$ colours and galaxy ages is not actually linear (see the curvature in the age trend in the top right panel of Fig.~\ref{fig:UVJ_figure}).  Thus, we do not give the functional form of our linear fit as we do not believe that it accurately represents a quantitative relationship between $UVJ$ colours and galaxy ages.  We also apply multiple linear and polynomial regression algorithms to attempt to derive quantitative relationships between $UVJ$ colours and our metallicities and ages (not shown).  Specifically, we attempt to find the best linear or polynomial combination of $U-V$ and $V-J$ colours that minimizes the residuals in age and metallicity, similar to the $S_Q$ axis defined in \cite{Belli_2019}.  However, we find that no simple functional form fits the data well and captures the relationships between $UVJ$ colours and ages and metallicities.       

Accordingly, we suggest that the $UVJ$ diagram can, as found previously, be used to identify whether a quiescent galaxy is relatively older or younger, but it may not be possible to specify a precise value of age based on an individual galaxy's $UVJ$ colours.  Additionally, we show that the $UVJ$ diagram can also be used to estimate whether a quiescent galaxy is relatively more metal-rich or metal-poor, though the scatter in metallicity is larger than in age (and again, it is difficult to specify an exact metallicity based on $UVJ$ colours alone, see also \citealt{Leja_2019_uvj}).  Other fitting degeneracies discussed in Section~\ref{sec:introduction} could be complicating these relationships, for example degeneracies between age, dust, and SFH (see Section~\ref{sec:model_implications} for further discussion).  Looking forward, more advanced fitting techniques, such as random forest algorithms or neural networks, in addition to combining photometry and spectra and developing SPS models further, may be able to provide a quantitative translation between $UVJ$ colours and ages and metallicities for individual galaxies.  

\subsection{Implications for galaxy evolution studies}\label{sec:science_implications}
In Section~\ref{sec:tracks}, we examined how trends between our spectroscopically measured ages and metallicities and independent photometric $UVJ$ colours compare to predictions from popular SPS models.  We found that, while the age trend is generally reproduced well by both \textsc{fsps} and BC03, these models do not recover the metallicity trend.  Here, we discuss the implications of our results for galaxy evolution studies in general.

\begin{figure*}
    \centering
    \includegraphics[width=\textwidth]{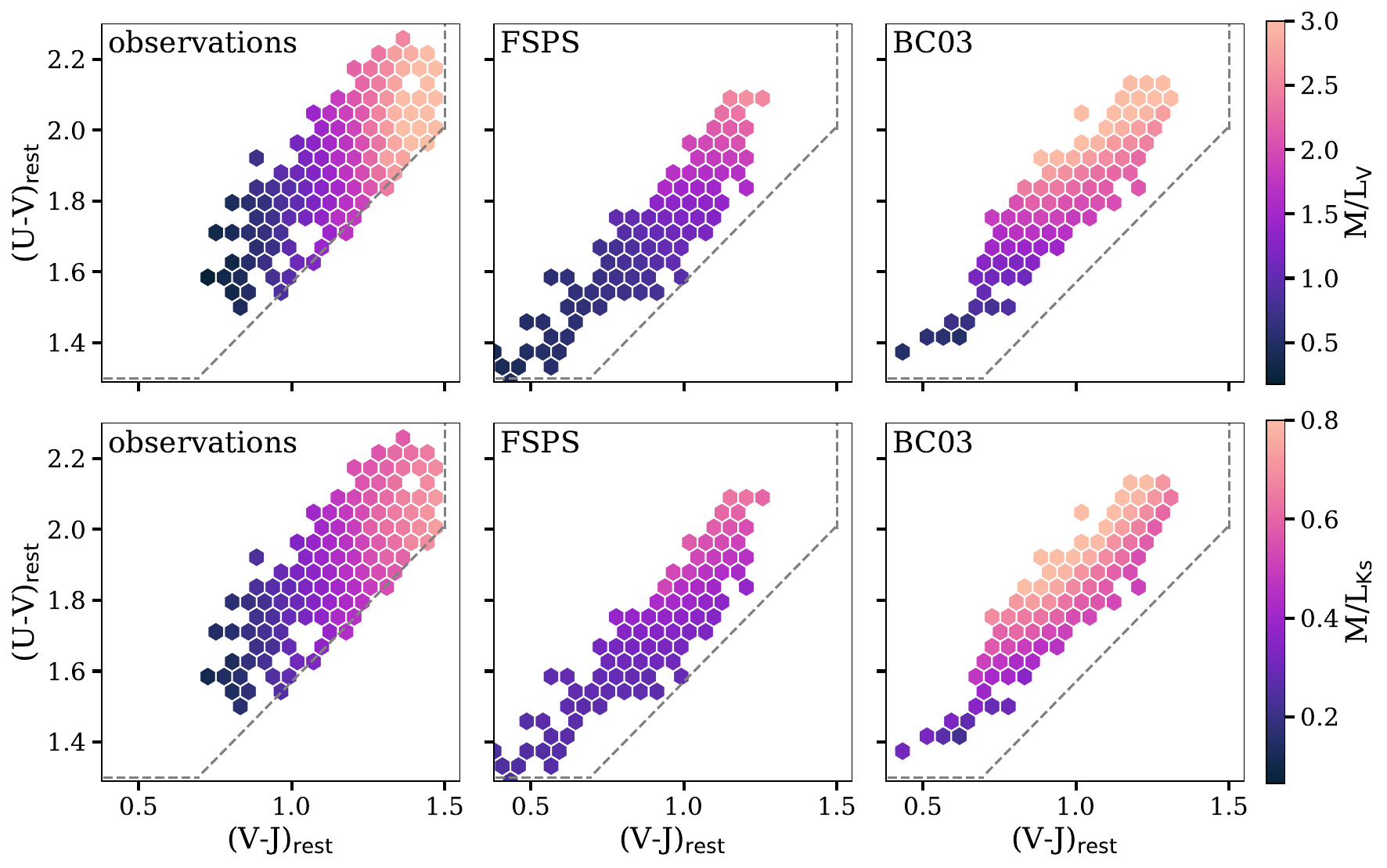}
    \caption{LOESS-smoothed rest-frame $UVJ$ diagrams for the data and the models, where the diagrams are colour coded by the $M/L$ ratio.  The LEGA-C $UVJ$ diagrams are shown in the left column, the \textsc{fsps} diagrams are shown in the middle column, and the BC03 diagrams are shown in the right column.  This is similar to Fig.~\ref{fig:UVJ_model_test}, where we input the ages and metallicities of each galaxy in our LEGA-C sample into the models, and output the predicted rest-frame $UVJ$ colours (from which we compute luminosities) and stellar masses. For the LEGA-C sample, we compute $M/L$ ratios using independently measured, dynamically scaled virial masses from \citet{van_der_wel_2022}.  The top row is colour coded by $M/L$ ratios in the $V$ band ($M/L_{\mathrm{V}}$) and the bottom row is colour coded by $M/L$ ratios in the $K_{\mathrm{s}}$ band ($M/L_{\mathrm{K}_\mathrm{s}}$).  While the $M/L$ ratio increases approximately along the quiescent sequence in most cases, the trend curves in opposite directions for the data and the models at the reddest tip of the quiescent sequence.  Additionally, the trends in the $M/L$ ratios predicted by the two models are not consistent with each other.}
    \label{fig:ML_vir}
\end{figure*}

SPS models have been used for decades to fit photometric and spectroscopic data of quiescent galaxies and interpret their formation and evolution.  However, if these models cannot fully reproduce observations as we suggest here, we may need to re-evaluate certain aspects of our understanding of galaxy evolution.  For example, while the age trends that we find in this work are qualitatively consistent with those predicted by the models, individual age measurements may not be reliable due to degeneracies with other fitting parameters \citep{Worthey_1994, Bell_deJong_2001, Papovich_2001, Bruzual_2003, Gallazzi_2005, Leja_2017, Leja_2019_sfh, Leja_2019_prospector}.  In particular, an incorrect metallicity will be compensated by an incorrect estimation of age, dust, or SFH.  Accordingly, derived metallicities, SFHs, and dust components may be systematically biased \citep{Leja_2017, Carnall_2019, Leja_2019_sfh, Leja_2019_prospector, Pacifici_2023, Bellstedt_2024}.  Thus, our work may have significant implications for our understanding of the SFHs of massive galaxies in the early Universe as derived from SED modelling (e.g., \citealt{Glazebrook_2017, Glazebrook_2024, Carnall_2023, Carnall_2024, de_Graaff_2024, Bevacqua_2025, Nanayakkara_2024}), or for why massive quiescent galaxies are still quite dusty (e.g., \citealt{Gobat_2018, Lee_2024_dustyQ, Setton_2024, Siegel_2024}). 

Another set of key quantities that could be affected is the $M/L$ ratios and therefore stellar masses.  To explore this issue, we compare trends between $M/L$ ratios in our data and those predicted by the models.  It has been found previously that $M/L$ ratios, similar to ages, have a tight relationship with rest-frame colours (e.g., \citealt{Bell_deJong_2001, van_de_sande_2015}).  Thus, in Fig.~\ref{fig:ML_vir} we show a similar comparison between our observations and SPS models as in Fig.~\ref{fig:UVJ_model_test}, but now colour coding the rest-frame $UVJ$ diagrams with $M/L$ ratios.  

In the top row we show the $V$-band $M/L$ ratios ($M/L_{\rm V}$) and in the bottom row we show the $K_{\rm s}$-band $M/L$ ratios ($M/L_{{\rm K}_{\rm s}}$).  We show the LEGA-C galaxies in the left panels, where we compute $M/L$ ratios using dynamical-model-calibrated virial masses derived in \cite{van_der_wel_2022}.  We derive luminosities from the \textsc{eazy} $V$- and $K_{\rm s}$-band colours (redetermined as in Section~\ref{sec:data_sample}, using the LEGA-C spectroscopic redshifts).  The virial masses were scaled to dynamical masses obtained via Jeans modelling in \cite{van_Houdt_2021}, and are defined as
\begin{equation}
    M_{\rm vir} = K(n)K(q)\frac{\sigma^{'2}_{*,{\rm int}}R_{\rm sma}}{G}
\end{equation}
(equation (9) in \citealt{van_der_wel_2022}), where $K(n) =  8.87 - 0.831n + 0.0241n
^2$ and where $n$ is the S\'ersic index \citep{Cappellari_2013}, $K(q) =  (0.87 + 0.38e^{-3.78(1 - q)})^2$ and where $q$ is the projected axis ratio $q = b/a$, $\sigma^{'2}_{*,{\rm int}}$ is the spatially integrated stellar velocity second moment from LEGA-C, and $R_{\rm sma}$ is the semimajor axis of the ellipse that contains 50 per cent of the projected S\'ersic light model from the best-fitting Jeans model.  

We use dynamical masses instead of stellar masses as they are independent of SPS models (see also \citealt{van_de_sande_2015}), similar to our age and metallicity measurements in the left panels of Fig.~\ref{fig:UVJ_model_test}.  As a caveat, we note that they include both baryonic and dark matter contributions\footnote{As these are quiescent galaxies, we neglect the gas mass.}.  The dark matter component may systematically bias the $M/L$ ratios to higher values; however, the qualitative trends along the $UVJ$ diagram would be the same assuming a similar dark matter fraction within one $R_{\rm e}$.  

In the middle and right panels of Fig.~\ref{fig:ML_vir}, we show the stellar $M/L$ ratios for the \textsc{fsps} and BC03 models, respectively, derived for each galaxy using the assumed age and metallicity (with added random scatter for the $UVJ$ colours as in Section~\ref{sec:tracks}) from which we compute luminosities.  As an additional caveat, we note that our treatment of dust could affect the observed luminosities while the assumed IMF will affect the model $M/L$ ratios.  We discuss this further in Section~\ref{sec:caveats}.

Similar to the metallicity panels in Fig.~\ref{fig:UVJ_model_test}, the qualitative trends between $M/L$ ratios in the data and in the models are not consistent.  In particular, while the $M/L$ ratio increases roughly along the quiescent sequence in most panels, the data and models curve in opposite directions at the reddest tip of the quiescent sequence.  Moreover, in the bottom-right panel, the $M/L_{{\rm K}_{\rm s}}$ ratio increases approximately perpendicular to the quiescent sequence in the BC03 models, while this is not the case for the data.  Thus, the data-model inconsistencies that we find in Section~\ref{sec:tracks} also apply to stellar mass measurements.  

Our results are consistent with those of \cite{van_de_sande_2015}, who compared $M/L$--colour relations to those predicted by SPS models for massive quiescent galaxies out to $z\sim2$.  They found that the BC03, \textsc{fsps}, and \cite{Maraston_2011} SPS models were not able to simultaneously reproduce the relationships that they found between dynamical $g$-band $M/L$ ratios and $(g-z)$ rest-frame colours and dynamical $K$-band $M/L$ ratios and $(g-K)$ rest-frame colour.  Furthermore, \cite{Bellstedt_2024} explored the impacts of different stellar libraries and model assumptions on SPS-model-derived stellar masses and found that the primary impact on stellar masses was in assuming a fixed, solar metallicity, followed by stellar library differences (see also \citealt{Jones_2022}) and SFH paramaterizations (see also \citealt{Carnall_2019, Leja_2019_sfh, Lower_2020, Suess_2022, Haskell_2024}).  

Additionally, the middle and right panels in Fig.~\ref{fig:ML_vir} demonstrate that there are significant differences in the $M/L$ ratio trends between the models themselves.  This was also indicated by \cite{van_de_sande_2015}, who found that introducing a variable IMF improved the match between observed and predicted $M/L$--colour relations for certain models, but not all (see also \citealt{Bellstedt_2024}).  Similarly, \cite{Muzzin_2009} found that different SPS models result in best-fitting stellar masses that differ by a factor of up to 1.5.  Finally, \cite{Lee_2025} found systematic differences in $M/L$ ratios measured using four different SPS models (\citealt{Vazdekis_2016}, BC03, \citealt{Charlot_2019}, and \textsc{fsps}). 

Given these findings, it would be informative to understand whether ages, metallicities, or other quantities are primarily driving these $M/L$ ratio differences.  To this end, we examine relationships between our measured ages and metallicities and $M/L$ ratios (not shown), and find that there may be a correlation between $M/L$ and age in our observations, albeit with large scatter.  There is no clear relation between $M/L$ and metallicity in our observed LEGA-C galaxies.  Additionally, we examine the ratio between the observed and modelled $M/L$ ratios (not shown).  In this case, there is also no clear relation with age or metallicity, indicating that we do not observe any specific biases for specific properties.  While this is unexpected given the inconsistencies between the models and the data in $UVJ$ space, it is not surprising that this analysis does not yield any further insights as the observed $M/L$ ratios derived from dynamical masses rely on many assumptions, have significant uncertainties, and include dark matter \citep{van_Houdt_2021, van_der_wel_2022}.  Thus, understanding the driver of $M/L$ ratio differences between the data and the models will benefit from a more in-depth investigation.   

\subsection{Implications for SPS models}\label{sec:model_implications}
In Section~\ref{sec:tracks}, we showed that popular SPS models such as \textsc{fsps} and BC03 can broadly reproduce the age trend in the $UVJ$ diagram, but cannot reproduce the [Fe/H] trend.  We also showed that these models fail to predict the observed $M/L$ ratio trends in Section~\ref{sec:science_implications}.  In this section, we discuss possible explanations for these data-model inconsistencies.

First, the observed discrepancies could be attributed to differences in the models used to measure stellar population parameters from absorption lines (i.e., \textsc{alf}) and the models used to produce the $UVJ$ colours (i.e., \textsc{fsps}, BC03).  While both \textsc{fsps} and \textsc{alf} use the same MIST isochrones \citep{MIST} and MILES stellar libraries \citep{Sanchez_Blazquez_2006}, the major difference between the SPS models considered here and \textsc{alf} is that \textsc{alf} allows for a variable abundance pattern while \textsc{fsps} and BC03 assume solar-scaled abundance patterns.  If models with $\alpha$-enhancements or variable abundance patterns predict different continuum shapes compared to models with solar-scaled abundance patterns, this could explain the differences between the observations and model predictions in Figs~\ref{fig:UVJ_model_test} and \ref{fig:ML_vir}.  For example, \cite{Choi_2019} found that the variable abundance pattern implemented in \textsc{alf} was necessary to simultaneously reproduce spectra and broad-band, rest-frame, optical colours ($ugr$).  \cite{Vazdekis_2015} found that $\alpha$-enhancement and a bottom-heavy IMF are required to reproduce observed photometric colours.  Several other groups have also introduced $\alpha$-enhanced models (e.g., \citealt{Coelho_2007, Lee_2009, Percival_2009, Knowles_2021, Knowles_2023, alpha_MC, alpha_BPASS}).  Thus, implementing more widespread use of non-solar abundance patterns, especially when studying $\alpha$-enhanced stellar populations such as those in distant massive quiescent galaxies, may be a promising way forward in explaining the discrepancies in $UVJ$ space.  In fact, \cite{alpha_MC} recently\footnote{At the time of writing, the $\alpha$-enhanced \textsc{fsps} models from \cite{alpha_MC} were not available as part of \textsc{python-fsps} for us to test here.} presented $\alpha$-\textsc{MC}, a new $\alpha$-enhanced implementation of the \textsc{fsps} models.  In the future, it will be interesting to generate the models in Section~\ref{sec:tracks} with $\alpha$-\textsc{MC} to examine how $\alpha$-enhancement affects the observed trends in stellar properties. 

In this context, it is interesting to note that these different approaches (solar-scaled versus variable abundance patterns) lead to different fitting results.  For example, \cite{Nersesian_2024} computed ages and metallicities for the LEGA-C galaxies using \textsc{prospector} (\citealt{prospector}, which makes use of the \textsc{fsps} models, \citealt{FSPS1, FSPS2}).  We find that there is good agreement between the \textsc{prospector} and \textsc{alf} ages (albeit with a large amount of scatter), but poor agreement between the metallicities (not shown, A. Nersesian, private communication).  This was also found in \cite{Beverage_suspense}, who fit the spectra of massive quiescent galaxies at $1 < z < 3$ with a \textsc{python} version of \textsc{alf} (\textsc{alf$\alpha$}, \citealt{alfalpha, Beverage_suspense}) and compared their ages and metallicities to \textsc{prospector} results from \cite{Slob_2024}.  Thus, implementing non-solar abundance patterns in standard fitting codes may improve the consistency in age and metallicity measurements between the different codes as well.  

Another potential reason for the discrepancies between the models and the data (from a modelling perspective) is the contribution of evolved stellar stages.  Specifically, the TP-AGB phase is prominent at low metallicities for young galaxies ($< 2$ Gyr) and can contribute significantly to the $M/L$ ratio \citep{Walcher_2011}.  The TP-AGB phase is implemented differently in every SPS model, and its exact contribution is debated (compare \citealt{Maraston_2005}, \citealt{Maraston_2006}, and \citealt{Lu_2024}, who found that a prominent TP-AGB phase reproduced observations the best, with \cite{Kriek_2010} who found that a prominent TP-AGB phase overpredicted near-infrared luminosities, and see \cite{Bevacqua_2025} as well).  These differences can have varying impacts on SED fits.  In particular, \cite{Muzzin_2009} fit the near-infrared SEDs of young galaxies, where the effects of the TP-AGB should be most evident, with different SPS models, and found that the best-fitting stellar population parameters differed by a factor of up to 3.  Thus, agreement on the implementation of TP-AGB stars and other evolved stellar stages is needed. 

Other aspects of SPS models could help alleviate the discrepancy with observations.  For instance, it will also be important to reconcile different spectral libraries, as they have been found to produce different results.  In particular, \cite{Byrne_2023} compared multiple spectral libraries using the same SPS model and found that the discrepancies with spectroscopic quantities when using different libraries were worse for metallicities than for ages, as the $4000$ \AA\ break was relatively insensitive to library differences.  Similar results were found by \cite{Bellstedt_2024}.  We note that \textsc{alf} makes use of the C3K stellar library, which \cite{Byrne_2023} found to be the most robust out of their tested libraries.  Finally, different SPS models do not always agree on the definition of metallicity.  For example, total metallicity is often computed using a combination of [Mg/Fe] and [Fe/H] (e.g. \citealt{Thomas_2003}) when Mg and Fe are both measured individually.  However, we note that the opacities are primarily driven by iron.  Moreover, as discussed above, both \textsc{fsps} and BC03 assume solar-scaled abundances, while massive quiescent galaxies are $\alpha$-enhanced.  In particular, \cite{Beverage_suspense} showed that the best-fitting \textsc{fsps} metallicities follow [Fe/H], and not [Z/H], for galaxies that are $\alpha$-enhanced (like the ones in this sample).  See also \cite{Choi_2014} and \cite{Leja_2019_sfh}.    

In summary, our results demonstrate that standard SPS models do not predict the correct $UVJ$ colours for quiescent galaxies when given stellar metallicities or ages (even when taking into account dust and extended SFHs, see Section~\ref{sec:caveats}).  Thus, we emphasize the need for caution when interpreting or assuming metallicities or stellar masses predicted from photometry, or derived via solar-abundance-scaled codes that rely on the shape of the continuum (see also \citealt{Bellstedt_2024, Nersesian_2024, Beverage_suspense}).  More work is needed to implement non-solar abundance patterns (see e.g. \citealt{alpha_MC}), understand the contribution of continuum shapes and advanced stellar stages, and reconcile the ingredients in different SPS models.  

\subsection{Caveats}\label{sec:caveats}
Here, we discuss the main caveats that should be taken into consideration when interpreting our results, including whether our observed trends may be driven by other physical properties or assumptions that we have made, and our use of \textsc{alf}.  

Throughout this work, we assume that there is no dust in the LEGA-C galaxies and we do not implement a dust law in our generation of \textsc{fsps} and BC03 models.  However, it is likely that at least some of the LEGA-C galaxies are affected by dust.  In particular, the location of our data in the $UVJ$ diagram in Figs~\ref{fig:loess_tracks}, \ref{fig:UVJ_model_test}, and \ref{fig:ML_vir} is slightly redder than expected from the models, especially when we compare with \textsc{fsps}, and dust could be contributing to this offset.  The presence of dust would also introduce additional scatter in the SPS model panels (middle and right) in Figs~\ref{fig:UVJ_model_test} and \ref{fig:ML_vir}.  We test these possibilities by generating the same models as shown in Figs~\ref{fig:UVJ_model_test} and \ref{fig:ML_vir}, but assuming a \cite{Calzetti_2000} dust law with (i) \textsc{fast} \citep{FAST1, FAST2} $A_V$ values, measured by and obtained via private communication with the LEGA-C team for each galaxy (although these values will be heavily affected by the degeneracies and model inaccuracies that we discuss throughout the paper), and (ii) $A_V$ values that scale with age (as dust approximately follows the age trend in the $UVJ$ diagram, see the arrow in Fig.~\ref{fig:loess_tracks}), such that $A_V = 0.5$ at 1 Gyr, decreasing down to an age of 3 Gyr, beyond which $A_V = 0$.  We correct the model \textsc{fsps} and BC03 luminosities for the respective dust extinction in each case (not shown).  In all cases, dust moves the distribution along the arrow shown in Fig.~\ref{fig:loess_tracks}, but does not provide a better match between the observations and the models.  However, we note that implementing the \textsc{fast} $A_V$ values for each galaxy broadens the distribution of the models in $UVJ$ space in the middle and right panels of Figs~\ref{fig:UVJ_model_test} and \ref{fig:ML_vir} to better match the observations.  Dust also acts to increase the $M/L$ ratios; however, it does not change their distribution in $UVJ$ space, preserving the discrepancy between the observations and the models.  While we find that dust does not drive the disagreement between the observations and the models, we note that correlations between galaxy properties could alter the distribution of galaxies in $UVJ$ space.  Thus, it is still possible that better $A_V$ constraints could bring the observations and models into agreement.  Further exploration of methods to disentangle dust from other physical properties is required.

Additionally, in our \textsc{fsps} and BC03 model generation, we assume a relatively simple SFH with a short, truncated burst.  We also assume a single age in our \textsc{alf} fits.  However, if multiple stellar populations are present (see e.g. \citealt{Carnall_2024, Nanayakkara_2024}), this could also play a role in the offset that we see between the data and the models.  We test the effects of several different SFHs, by extending the burst, varying the $\tau$ value, and implementing a delayed-$\tau$ model.  We find that altering the assumed SFH may help to explain some of the scatter that we see in the right-most panels of Figs~\ref{fig:UVJ_model_test} and \ref{fig:ML_vir}, but does not account for all of it.  Crucially, implementing different SFHs does not generate the trends that we observe in age, metallicity, or $M/L$ ratios, with these parameters behaving the same way in these test models as with our assumed simple SFH. 

There may be other effects creating the trends in stellar population parameters that we see that are not inherent to the SPS models.  For example, the age and metallicity measurements from which we derive our main results are based solely on our fits using \textsc{alf}, which may not be as accurate for galaxies near the edge of the $UVJ$ box; these galaxies could potentially be dusty post-starburst galaxies \citep{Zick_2018} or have low levels of star formation \citep{Akins_2022}.  Since our conclusion of a metallicity trend depends on this region of the $UVJ$ diagram, we test shifting the $UVJ$ box by 0.1 magnitudes toward redder $U-V$ colours to examine the effect of excluding potential contaminant galaxies near the edge of the $UVJ$ box.  We find that both the age and metallicity trends are still present and follow the same qualitative direction as in Fig.~\ref{fig:UVJ_figure}, although the age trend is slightly tighter.  Thus, the inclusion of the galaxies towards the edge of the $UVJ$ box does not affect our conclusions. 

To quantitatively examine the potential biases of using \textsc{alf}, we additionally test whether \textsc{alf} could be predicting lower [Fe/H] abundances for a quiescent population which had a recent minor starburst.  In this case, the equivalent widths of absorption lines would be decreased by an additional, mostly featureless, continuum contribution.  To test this scenario, we select 33 quiescent galaxies from LEGA-C that lie along the quiescent sequence, but away from the edge of the $UVJ$ box.  For each selected LEGA-C galaxy, we generate a young ($0.1$ Gyr) SED with \textsc{fsps} that is 6 per cent of the luminosity of the corresponding LEGA-C galaxy in the $V$ band, and add the young SED to the LEGA-C spectrum to make 33 composite spectra that still look visually quiescent.  We determine this luminosity percentage by estimating the $UVJ$ colours of the composite spectra such that the composite galaxies lie near the edge of the $UVJ$ box.  Note that we give the young SEDs the same metallicity as derived from the fiducial fit to the corresponding LEGA-C spectrum.  We also test giving the young SEDs higher and lower metallicities and find that this does not affect our conclusions.  We fit the composite spectra with \textsc{alf} in the same way as the corresponding LEGA-C spectra.  We find that including a young population on top of an older quiescent one results in [Fe/H] values that are consistent with the [Fe/H] values that we derive from the fiducial fits within uncertainties.  However, we note that in many cases the fits are driven to slightly lower median [Fe/H] values compared to the fiducial case.  In particular, we find that an average decrease of $\sim 0.24$ in $U-V$ colour and $\sim 0.04$ in $V-J$ colour results in an average decrease in [Fe/H] of $\sim 0.05$ dex.  When we compare this to our expectations from a similar decrease in $U-V$ and $V-J$ in our observations in the lower right panel of Fig.~\ref{fig:UVJ_figure}, we see that the average observed decrease in [Fe/H] is $\sim 0.11$ dex.  Therefore, low levels of star formation may contribute to the observed metallicity trend, and in general we caution that composite stellar populations near the edge of the $UVJ$ box may lead to lower metallicities.  However, based on the test presented here, these low levels of star formation cannot fully explain the observed metallicity trend, and we note that such a recent burst would generate emission lines, which we do not observe in our spectra.

In the future, it will still be important to confirm our \textsc{alf} fitting results via other means.  Encouragingly, we note that our observed metallicity trend is consistent with the results of \cite{Nersesian_2025}, who simultaneously fit photometry and spectroscopy in a large sample of star-forming and quiescent LEGA-C galaxies.  Another possibility is to estimate ages and metallicities via traditional Lick indices \citep{Worthey_lick}, which could potentially provide us with insight into whether our results are robust.  However, these measurements would not have the same independence as our \textsc{alf} measurements, as SPS models like \textsc{fsps} or BC03 are required to interpret Lick indices.  Additionally, this method will likely not be able to provide age and metallicity estimates to the same level of accuracy as \textsc{alf} as it only makes use of a few individual features while \textsc{alf} considers the full spectrum (see e.g. \citealt{Kriek_2016}).  Other methods of constraining ages and metallicities are beyond the scope of this work, but should be considered in the future. 

Finally, we test whether the metallicity trend could be caused by distinct types of galaxies populating different parts of the $UVJ$ diagram.  For example, if a specific population with low metallicity is preferentially found towards the edge of the $UVJ$ box, this could drive the observed trend.  For quiescent galaxies, properties that are typically correlated with lower metallicity include lower velocity dispersion, lower mass, or higher redshift (e.g., \citealt{Smith_2009, Leethochawalit_2019, Beverage_2023, HM_Beverage, Beverage_suspense}).  We do not find any preference of lower velocity dispersion or lower mass populations towards the edge of the $UVJ$ box.  There is a slight preference of lower metallicities for higher redshift galaxies, which are indeed closer to the edge of the $UVJ$ box.  It is still unclear why these galaxies fall in this region of the $UVJ$ diagram; however, we note that our sample also contains fewer galaxies at these redshifts.  Larger samples of high redshift galaxies will be required going forward to examine whether there is a true dependence on redshift.  The only additional property that is preferentially found towards the edge of the box is slightly higher sSFR.  However, at the same mass, we do not expect quiescent galaxies with higher sSFR to have lower metallicity.  In fact, we would expect the opposite trend, as galaxies that quench at later times and have longer star-formation time-scales will have higher [Fe/H] values (e.g., \citealt{Beverage_2021, HM_Beverage, Beverage_suspense}).

\section{Summary and conclusions}\label{sec:conclusions}
The age--metallicity degeneracy \citep{Worthey_1994, Bruzual_2003, Gallazzi_2005} poses a major challenge to interpreting photometric data of quiescent galaxies.  Measuring metallicities has historically been challenging, especially beyond the local Universe, where low S/N data make it difficult to constrain weak metallicity-sensitive features.  As a result, metallicity has typically been left unconstrained or fixed to solar values, leading to (systematic) uncertainties in stellar population studies.  Consequently, the red colours of quiescent galaxies, for example in the well-known $UVJ$ diagram, were fully attributed to age (e.g., \citealt{Whitaker_2012, Whitaker_2013, Mendel_2015, Belli_2019, Diaz_Garcia_2019}), and our knowledge of how metallicity affected the shapes of galaxy continua was completely dependent on solar-scaled models, which become increasingly uncertain at earlier times.  

In this work we compared, for the first time, spectroscopically derived ages and metallicities (from continuum-normalized, absorption-line spectra) to independently measured broad-band colours of quiescent galaxies beyond the low-redshift Universe.  In particular, we examined a sample of $\sim 700$ massive quiescent galaxies at $0.6\lesssim z \lesssim 1.0$ from the LEGA-C survey \citep{van_der_Wel_2021}.  We fit the continuum-normalized, absorption-line spectra using \textsc{alf} \citep{CvD_2012a, Conroy_2018} and recovered SSP-equivalent ages and [Fe/H] values.  We compared these measurements to UltraVISTA \citep{McCracken_2012} $UVJ$ colours \citep{Muzzin_2013} by colour coding the $UVJ$ diagrams of these galaxies by age and metallicity.  We also compared our observed trends between spectroscopic ages and metallicities and broad-band $UVJ$ colours to predictions from popular SPS models including \textsc{fsps} \citep{FSPS1, FSPS2} and BC03 \citep{Bruzual_2003}.  Our findings can be summarized as follows:
\begin{itemize}
    \item Galaxy age generally increases along the quiescent sequence in the $UVJ$ diagram.  This was implied in \cite{Whitaker_2012, Whitaker_2013}, \cite{Mendel_2015}, \cite{Belli_2019}, and \cite{Diaz_Garcia_2019}; however, for the first time, we detected this trend using age measurements that are independent of $UVJ$ colours, while also accounting for stellar metallicity.  We also found that there is a slight curvature to the trend, with the oldest galaxies populating the reddest end of the $U-V$ axis, but located slightly blueward on the $V-J$ axis.  Thus, we concluded that a simple linear relation between $UVJ$ colours and ages does not represent the age trend very well at these redshifts, in contrast to the findings of \cite{Belli_2019} at $1.0 < z < 2.5$.
    \item Stellar metallicity increases approximately perpendicular to the quiescent sequence.  \textit{Thus, the sequence in metallicity does not follow the sequence in age}, in contrast to predictions from stellar population models \citep{FSPS1, FSPS2, Bruzual_2003}.  This finding indicates that the interpretation of the quiescent sequence as an age sequence \citep{Whitaker_2012, Whitaker_2013, Mendel_2015, Belli_2019, Diaz_Garcia_2019} is robust.  
    \item In general, SPS models are able to qualitatively reproduce our observed age trend, though the observed galaxies are on average redder than model predictions.  This difference could be explained by some dust attenuation. 
    \item However, the models predict that metallicity increases nearly perpendicular to the trend that we observe in the data.  In other words, SPS models predict very different $UVJ$ colours for a given metallicity than the colours that we observe.  
    \item These data-model inconsistencies likely impact our understanding of stellar population properties of quiescent galaxies derived from photometric data, as an incorrect metallicity will be compensated by age, dust, or SFH, which may subsequently bias $M/L$ ratios.  To explore this further, we examined the trends between $M/L$ ratios and the $UVJ$ diagram measured in our observed LEGA-C galaxies and predicted by \textsc{fsps} and BC03.  For our observations, we made use of independent $M/L$ ratios derived from dynamically scaled virial masses.  Similar to the metallicity trend, we found that the $M/L$ ratio trends predicted by SPS models behave very differently from those which we observe.  This difference indicates that stellar masses inferred from $M/L$ ratios measured via SPS models may also suffer from model inaccuracies.
\end{itemize}

Our work suggests that, while state-of-the-art SPS models are powerful tools in deriving stellar population properties, further development is required to accurately recover the stellar metallicities, ages, and $M/L$ ratios of quiescent galaxies.  In particular, the implementation of non-solar abundance patterns and a better understanding of the contribution from evolved stellar stages may improve the mismatch between observations and SPS model predictions, as well as the discrepancies between different SPS models.  The need to improve model predictions will only become more urgent as we continue to push the frontier of quiescent galaxy studies to higher redshifts with \textit{JWST}.

\section*{Acknowledgements}
We thank the anonymous referee for taking the time to give us useful feedback that improved this manuscript.  We thank the LEGA-C team for making their data set public.  We also thank Brian Lorenz, Angelos Nersesian, and Colin Yip for useful conversations.  This work was performed using the compute resources from the Academic Leiden Interdisciplinary Cluster Environment (ALICE) provided by Leiden University.  This work also used the Dutch national e-infrastructure with the support of the Samenwerkende Universitaire Rekenfaciliteiten (SURF) Cooperative using grant no. EINF-6344 and grant no. EINF-10017 which are financed by the Dutch Research Council (NWO). MK acknowledges funding from the NWO through the award of the Vici grant VI.C.222.047 (project 2010007169) and NSF AAG grant AST-1909942.  PEMP acknowledges the support from the Dutch Research Council (NWO) through the Veni grant VI.Veni.222.364.

%%%%%%%%%%%%%%%%%%%%%%%%%%%%%%%%%%%%%%%%%%%%%%%%%%
\section*{Data Availability}
This study makes use of data from the LEGA-C survey.  The 2D and reduced 1D spectra can be obtained from the ESO Science Archive Facility (\url{http://archive.eso.org/eso/eso_archive_main.html}).  The reduced spectra and catalogue have been released by ESO \newline (\url{http://archive.eso.org/cms/eso-archive-news/Third-and-final-release-of-the-Large-Early-Galaxy\\-Census-LEGA-C-Spectroscopic-Public-Survey-published.html}) and are also available here: \url{https://users.ugent.be/~avdrwel/research.html\#legac}.  For details, see \cite{van_der_Wel_2016, van_der_Wel_2021} and \cite{Straatman_2018}.

The re-determined $UVJ$ colours as well as ages, metallicities, and [Mg/Fe] values are available in an online catalogue.  Other data products generated in the course of this work will be made available upon reasonable request. 

\bibliographystyle{mnras}
\bibliography{references} % if your bibtex file is called example.bib

%%%%%%%%%%%%%%%%%%%%%%%%%%%%%%%%%%%%%%%%%%%%%%%%%%

%%%%%%%%%%%%%%%%% APPENDICES %%%%%%%%%%%%%%%%%%%%%

%\appendix

%%%%%%%%%%%%%%%%%%%%%%%%%%%%%%%%%%%%%%%%%%%%%%%%%%

% Don't change these lines
\bsp	% typesetting comment
\label{lastpage}
\end{document}